\documentclass[%
 reprint,
 amsmath,amssymb,
 aps,
 onecolumn,
11pt]{revtex4-2}

\usepackage{subcaption}
\usepackage{float}
\usepackage{enumitem}
\usepackage{comment}
\usepackage{color}
\usepackage{graphicx}
\usepackage{subeqnarray}
\usepackage{dcolumn}
\usepackage{bm}
\usepackage[normalem]{ulem}

\usepackage[
scale=0.75, marginratio={1:1, 1:1}, ignoreall,
]{geometry}

\newcommand{\p}{\partial}

\providecommand\bnabla{\boldsymbol{\nabla}}
\providecommand\bDelta{\boldsymbol{\Delta}}
\newcommand{\uu}{\textbf{u}}
\newcommand{\xx}{\textbf{x}}

\newcommand{\AL}[1]{\textcolor{red}{#1}} 
\begin{document}

\preprint{APS/123-QED}

\title{\Large{The effects of spanwise confinement \\ on stratified shear instabilities}}

\author{Yves-Marie Ducimeti\`ere}

 \email{yves-marie.ducimetiere@epfl.ch}
 \author{Fran\c cois Gallaire}

\affiliation{Laboratory of Fluid Mechanics and Instabilities, EPFL, CH1015 Lausanne, Switzerland}%


\author{Adrien Lefauve }

\author{Colm-cille P. Caulfield}

\altaffiliation[Also at ]{BP Institute, University of Cambridge, Madingley Road, Cambridge CB3 0EZ, UK}

\affiliation{Department of Applied Mathematics and Theoretical Physics, University of Cambridge, \\ Centre for Mathematical Sciences, Wilberforce Road, Cambridge CB3 0WA, UK %
\vspace{0.5cm}}%


\date{\today}

\begin{abstract}
We consider the influence of transverse confinement on the  instability properties of velocity and density distributions reminiscent of those pertaining to exchange flows in stratified inclined ducts, such as the recent experiment of Lefauve et al. (\textit{J. Fluid Mech.} \textbf{848}, 508-544, 2018). Using a normal mode streamwise and temporal expansion for flows in ducts with various aspect ratios $B$ and non-trivial transverse velocity profiles, we calculate two-dimensional (2D)  dispersion relations with associated eigenfunctions varying in  the `crosswise' direction, in which the density varies, and the spanwise direction, both  normal to the duct walls and to the flow direction.  We also compare these 2D dispersion relations to the so-called one-dimensional (1D) dispersion relation s obtained for spanwise invariant perturbations, for different aspect ratios $B$ and bulk Richardson numbers $Ri_b$. In this limited parameter space, the presence of lateral walls has a stabilizing effect, in that the 1D growth-rate predictions are almost systematically an upper bound to the 2D growth-rates, which in turn decrease monotonically as lateral walls are brought together with increased spanwise confinement ($B\rightarrow 0$). 
Furthermore, accounting for spanwise-varying perturbations results in a plethora of unstable modes, the number of which increases as the aspect ratio is increased. These modes present an odd-even regularity in their spatial structures, which is rationalized by comparison to the so-called one-dimensional oblique (1D-O) dispersion relation obtained for oblique waves, characterized by a continuously varying spanwise wavenumber in addition to the streamwise wavenumber. Finally, we show that in most cases, the most unstable 2D mode is the one that oscillates the least in the spanwise direction, as a consequence of viscous damping. However, in a limited region of the parameter space and in the absence of stratification, we show that a secondary mode with a more complex `twisted' structure dominated by crosswise vorticity becomes more unstable than the least oscillating Kelvin-Helmholtz mode associated with spanwise vorticity.
\end{abstract}

\maketitle

\clearpage

\setcounter{tocdepth}{3}
\setlength\parindent{0pt} 
    \makeatletter
    \let\toc@pre\relax
    \let\toc@post\relax
    \makeatother

\clearpage

\clearpage

\section{Introduction}

Flows in the natural environment (such as in the atmosphere or ocean)  are often stably stratified in the vertical, with the horizontally-averaged density decreasing with height. Such environmental flows are also often characterised
by a background velocity distribution that decreases with height, 
resulting in vertical shear. This combined effect of buoyancy and shear results in a large variety of interesting dynamical behaviors exhibited by stratified shear flows. An important ingredient influencing such behaviors is the spatial confinement, inherent to many geophysical flows such as valleys, estuaries \cite{Geyer10}, submarine canyons, straits or deep ocean trenches.\\

Perhaps the most classical example of this dynamical behavior is the  overturning Kelvin-Helmholtz instability (perhaps more appropriately called a Rayleigh instability when the region 
of inflectional shear has a finite depth) as well as so-called Holmboe instabilities, typically associated with relatively `sharp' density gradients, which all contribute to  the mixing and transport of heat, salt or indeed various pollutants. In the Kelvin-Helmholtz instability, a single instability mode appears, traveling at the mean velocity of the fluid layer, which can grow into an array of elliptical vortical billows, that in turn overturn and smear out the density interface. In contrast, the Holmboe instability gives rise to propagating modes, which are associated (at finite amplitude) with vortices displaced from the density interface, which typically survives the ensuing scouring motion.\\

The Holmboe instability has attracted in recent years a large variety of numerical and experimental studies. Direct numerical simulations (DNS) have enabled a thorough description of nonlinear saturation and mixing mechanisms \cite{Smyth91,Smyth07,Carpenter07,Salehipour16,Salehipour2018,Smith21}. Laboratory experiments have been conducted in salt-stratified exchange flows that also investigate various aspects of this instability \cite{Caulfield95,Tedford09,Carpenter10,Meyer14}.  In particular, \cite{Lefauve18} investigates the laboratory-scale realization of the Holmboe instability. A sustained stratified shear flow is generated in the laboratory by exchange flow through an inclined square duct, connecting two reservoirs filled with fluids of different densities. The duct confines the flow in the `crosswise' direction (slightly tilted from the vertical) along which it is stratified, as well as in the spanwise direction.\\

In \cite{Lefauve18}, a three-dimensional, nonlinear and asymmetric Holmboe wave (in this context  the finite amplitude manifestation of the instability) was  observed and characterized in detail using three-dimensional, volumetric measurements
of the velocity and density fields. A temporal local linear stability analysis on the two-dimensional, streamwise-averaged, experimental flow was also performed. Three-dimensional perturbations were sought, having two-dimensional, cross-sectionally confined eigenfunctions and a streamwise normal mode expansion. The matching between  the resulting most unstable eigenmode developing on the mean flow and the experimental structure proved to be (perhaps surprisingly) excellent, validating the \textit{a posteriori} linear stability approach, where the time-averaged flow used captured the nonlinear effects
of the various perturbations. \newline  
 
Of great interest in \citep{Lefauve18}, is the importance of the spanwise confinement to the dynamical evolution of the Holmboe wave. With $v$ and $\omega$ designating the spanwise velocity and vorticity respectively, they observed that the  'presence of lateral walls gives rise to relatively large spanwise gradients $\left | \partial_y v \right |$, positive in the centre of the duct and negative near the boundaries . These gradients have a vortex stretching effect on $\omega_y$ [...] producing negative vorticity in the centre (reinforcing the mean shear), and positive vorticity near the boundaries (weakening the mean shear)' (p.534). As mentioned previously, such a  strong confinement effect should also be present in many geophysical flows. In this context,  \citep{Lefauve18} also draws attention to the fact that 'laboratory observations in confined geometries are often compared to stability analyses that ignore confinement, and numerical simulations usually impose periodic boundary conditions in the spanwise direction’ (p.540). The extent to which properties of three-dimensional confined (in the transverse, along-crest direction) waves are well predicted by such analyses is difficult to predict \textit{a priori}, and remains an open question. We aim to address this question here, considering a flow  configuration motivated by the experimental 
geometry described in \citep{Lefauve18}.  \newline

To address this aim, we will proceed as follows. In \ref{sec1}, we describe our linear $2D$ stability analysis and recall some fundamental results from the related literature. In \ref{sec2}, dispersion relations of the most unstable confined modes are presented in selected parameters space; they are compared to a classical $1D$ (in the sense of being spanwise invariant) predictions . In \ref{sec:harmo}, we restrict ourselves to a particular wavenumber and analyze in greater detail the unstable part of the spectrum. We  compare the associated predictions to generalized $1D$ predictions, allowing oblique modes. In \ref{sec:Bmode}, we analyze in details a specific mode that appears actually to be destabilized by the presence of transverse walls. Finally, we draw some relatively brief conclusions, and suggest potential further directions of research.
\section{Problem formulation, background and objectives} \label{sec1}
\subsection{Formulation}

\subsubsection{Flow configuration and notations}


\begin{figure}
\centering
  \centerline{\includegraphics[trim={4.0cm 18.5cm 3.5cm 3.5cm},clip,width=0.9\linewidth]{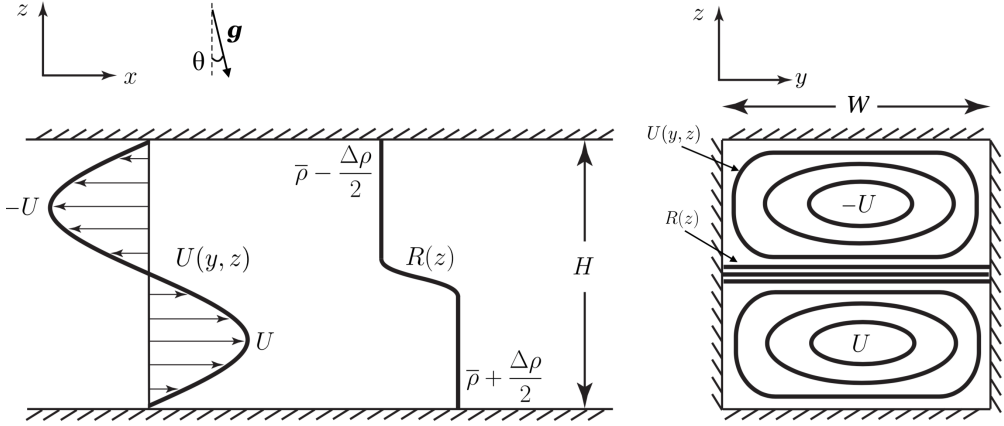}}
\caption{Schematic of our confined duct flow configuration (dimensional variables). }
\label{fig:sch3D}
\end{figure}

The flow configuration is illustrated in figure \ref{fig:sch3D}. The streamwise ($x$) axis is aligned along the duct, the spanwise ($y$) axis is across it, and the crosswise ($z$) axis is tilted at an angle $\theta$ from the true vertical, resulting in a nonzero projection of the gravity driving the exchange flow in the streamwise direction. The duct is assumed infinitely long in the streamwise direction to avoid end effects. The coordinate system is centered in the middle of the duct cross-section, such that $-H/2 \leq z \leq H/2$ and $-W/2 \leq y \leq W/2$, where $H$ is the duct crosswise `height' and $W$ the duct spanwise `width'. The velocity field is $\uu(x,y,z,t)=(u,v,w)$, the density field is $\rho(x,y,z,t)$ and the pressure field is $p(x,y,z,t)$. The base velocity profile $U(y,z)$ and density profile $\mathcal{R}(z)$ will be specified in section \ref{sec:bf}.\newline

To nondimensionalize the problem we choose to scale velocities by half the total (peak-to-peak) velocity jump in the base flow: $(\tilde{u},\tilde{v},\tilde{w}) = (u,v,w)/(\Delta U /2)$, and lengths by half the height of the duct: $(\tilde{x}, \tilde{y}, \tilde{z}) = (x,y,z)/(H /2)$, such that $-1 \leq \tilde{z} \leq 1$ and $-B \leq \tilde{y} \leq B$, where 
\begin{equation}
    B = \frac{W}{H}
\end{equation} 
is the duct aspect ratio; $B<1$ corresponds to what we refer to as a `narrow' duct, $B=1$ to a square duct, and $B>1$ corresponds to what we refer to as a wide duct. The corresponding nondimensional advective time is $\tilde{t} = t / (H/\Delta U)$. Finally, the nondimensional density   is $\tilde{\rho} = (\rho-\rho_0)/(\Delta \rho/2)$, where $\rho_0$ is the mean reference value and $\Delta \rho/2$ is half the (peak-to-peak) density jump in the density base profile.  

\subsubsection{Governing equations}

We model the flow by the incompressible Navier-Stokes equations under the Boussinesq approximation, which requires $\Delta \rho / \rho_0  \ll  1 $ (valid for the experimental flow configuration of \cite{Lefauve18}),  
i.e. the density difference only plays a role through the reduced gravity ${g}' = g \Delta \rho/\rho_0$. 
In addition, the kinematic viscosity ($\nu$) and the mass (salt) diffusivity ($\kappa_m$) are assumed constant. 
Dropping the tildes, we obtain the following set of nondimensional governing equations: 
\begin{subeqnarray}\label{eq-motion}
\bnabla \cdot \uu &=& 0, \slabel{eq-motion-1}\\ 
\p_t \uu + \uu \cdot \bnabla \uu &=& -\bnabla p +  Ri_b \left(-\cos\theta \mathbf{\hat{z}}  + \sin\theta \mathbf{\hat{x}}   \right) \, \rho \ 
+ Re^{-1} \, \bDelta \uu, \slabel{eq-motion-2}\\
\p_t \rho + \uu \cdot \bnabla \rho &=& (Re \, Sc)^{-1}\,  \bDelta \rho, \slabel{eq-motion-3}
\end{subeqnarray}
where the nondimensional parameters are: 
\begin{equation}
Re =\frac{\Delta U H}{4\nu}, \ \ \ \ \ \ \ \ \ \ \ Ri_b = \frac{g \Delta\rho H}{\rho_0 (\Delta U)^2 } = \frac{{g}'H}{(\Delta U)^2}, \ \ \ \ \ \ \ \ \ \ \ Sc = \frac{\nu}{\kappa_m}.
\label{eq:ND}
\end{equation}
The Reynolds number, $Re$, compares advective to  diffusive time scales for the flow. The bulk Richardson number, $Ri_b$, compares the potential energy of the flow to the shear-induced kinetic energy. 
The Schmidt number, $Sc$, compares the diffusivity of momentum to the mass diffusivity. \newline

In a linear stability analysis, we expand our variables as:
\begin{subeqnarray}\label{eq-perturb}
\uu(\xx,t) &=& \left(U(y,z),0,0 \right) + \epsilon \check{\uu}(\xx,t) , \quad \quad \left |\epsilon \right |  \ll 1 \slabel{eq-perturb-1}\\ 
p(\xx,t) &=&  P(y,z) + \epsilon \check{p}(\xx,t),  \slabel{eq-perturb-2}\\
\rho(\xx,t)  &=&  \mathcal{R}(z) + \epsilon \check{\rho}(\xx,t), \slabel{eq-perturb-3}
\end{subeqnarray}
\noindent where $\xx = (x,y,z)$, i.e. as a sum of a steady base flow and arbitrarily small perturbations. The base flow is assumed  parallel to the duct and invariant in the streamwise direction $x$. Expansions of Eqs.(\ref{eq-perturb}) are then plugged into Eqs.(\ref{eq-motion}), leading to linear equations for the perturbations at $O( \epsilon)$. The $x$ (and $t$) invariance of the base flow and infinite extent of the domain in $x$ allow us to express any perturbation $\check{f}$ as Fourier modes in $x$ and $t$:
\begin{equation}
\check{f} = \hat{f}(y,z)\exp{(\mbox{i}kx+\sigma t)} + c.c
\label{eq:fdecomp}
\end{equation}
Since we consider temporal instabilities in this paper, we set the wavenumber $k \in \mathbb{R}$ and $\sigma \in \mathbb{C} $, such that the real part of $\sigma$ represents the growth rate while its imaginary part represents the frequency. \newline

As explained in \cite{Lefauve18}, the dimensionality of the system (number of flow variables) can be reduced at the cost of increasing its differential order. In this paper, keeping the primitive variable formulation ($\hat{u},\hat{v},\hat{w},\hat{\rho},\hat{p}$) would be preferred in order to avoid fourth order derivatives that, with the discretization method presented later, would lead to poor matrix conditioning. However, this leads to problematic storage requirements. Consequently, we adopted the compromise proposed in \cite{Hu12}:  $\hat{u}$ was eliminated, leading to at most third order derivatives of $\hat{v},\hat{w}$. Rewriting the continuity equation as  $\hat{u}=ik^{-1}(\partial_y \hat{v} + \partial_z \hat{w})$ and plugging it into the $x$-momentum equation results in  the following generalized eigenvalue problem (dropping the inverted hats):
\begin{equation}
\setlength{\arraycolsep}{1pt}
\renewcommand{\arraystretch}{1}
\sigma\left[
\begin{array}{cccc}
  \mathcal{I}  \ \ &  \ \ & \ \ & \ \  \\
   \ \ &  \mathcal{I} \ \ & \ \ & \ \  \\
   \ \ &  \ \ & \mathcal{I} \ \ & \ \  \\
   \partial_y \ \ & \partial_z \ \ & \ \ &  \ \  \\
\end{array}  \right] 
\setlength{\arraycolsep}{1pt}
\renewcommand{\arraystretch}{1}
\left[
\begin{array}{c}
  v \\
  w \\
  \rho\\
  p \\  
\end{array}  \right]=
\setlength{\arraycolsep}{1pt}
\renewcommand{\arraystretch}{1}
\left[
\begin{array}{cccc}
  \mathcal{L}_v  \ \ &    \ \ &   \ \ & \mathcal{L}_{vp}  \\
   \ \ &  \mathcal{L}_w  \ \ & \mathcal{L}_{\rho w}  \ \ & \mathcal{L}_{wp}  \\ 
    \ \ &  \mathcal{L}_{w \rho}  \ \ & \mathcal{L}_{\rho}  \ \ &   \\ 
  \mathcal{L}_{pv}  \ \ &  \mathcal{L}_{pw}  \ \ & \mathcal{L}_{p\rho}  \ \ & \mathcal{L}_{pp}  \\ 
\end{array}  \right]
\setlength{\arraycolsep}{1pt}
\renewcommand{\arraystretch}{1}
\left[
\begin{array}{c}
  v \\
  w \\
  \rho\\
  p \\  
\end{array}  \right],
\label{GenEig_2D}
\end{equation}
where:
\begin{subeqnarray*}
\mathcal{L}_v & = &  -ikU + Re^{-1} \Delta \slabel{Lv} ,\\
\mathcal{L}_{vp} & = & - \partial_y ,\\
\mathcal{L}_{w} & = &  -ikU + Re^{-1}\Delta ,\\
\mathcal{L}_{w\rho} & = &  - Ri_b\cos(\theta), \\
\mathcal{L}_{wp} & = &  - \partial_z ,\\
\mathcal{L}_{\rho w} & = & -\partial_z \mathcal{R} ,\\
\mathcal{L}_{\rho} & = & -ikU + (Sc  Re)^{-1}\Delta ,\\
\mathcal{L}_{pv} & = & -ikU\partial_y  + ik\partial_y U + Re^{-1}(-k^2\partial_y +\partial_{yyy} + \partial_{zzy} ) ,\\
\mathcal{L}_{pw} & = &  -ikU\partial_z  + ik\partial_z U + Re^{-1}(-k^2\partial_z +\partial_{yyz} + \partial_{zzz} ), \\
\mathcal{L}_{p \rho} & = & -ikRi_b \sin(\theta) ,\\
\mathcal{L}_{pp} & = & -k^2 .\\ 
\end{subeqnarray*}
Note that $\Delta$ is the Laplacian operator in $x$ Fourier space $\Delta = (-k^2+\partial_{zz} +\partial_{yy})$, and $\mathcal{I}$ is the identity operator. \newline

The solid and impermeable duct walls were modeled by no slip boundary conditions for velocities and no mass flux for the density:
\begin{subeqnarray}
v =  w = \partial_y v = \partial_y \rho & = & 0 , \quad \mbox{for} \quad y=\pm B, \quad \forall z ,\slabel{BC_vel_y_1} \\
v = w = \partial_z w = \partial_z \rho & = & 0,  \quad \mbox{for} \quad  z=\pm 1, \quad \forall y ,\slabel{BC_vel_y_2}
\end{subeqnarray}
where the $\partial_y v = \partial_z w = 0$ conditions  result from the continuity equation ($iku=-\partial_y v - \partial_z w = 0$ at the walls). \newline

The equations were discretized by a custom-built two-dimensional Chebyshev pseudospectral method. Note that this method represents a nontrivial improvement in accuracy and speed upon the finite-difference method of \cite{Lefauve18}. More details on the discretization and the numerical solution are given in Appendix~\ref{appnm}. 

\subsubsection{Base flows \label{sec:bf}}

\begin{figure}
  \centerline{\includegraphics[trim={3.5cm 10cm 3.5cm 10cm},clip,width=0.5\linewidth]{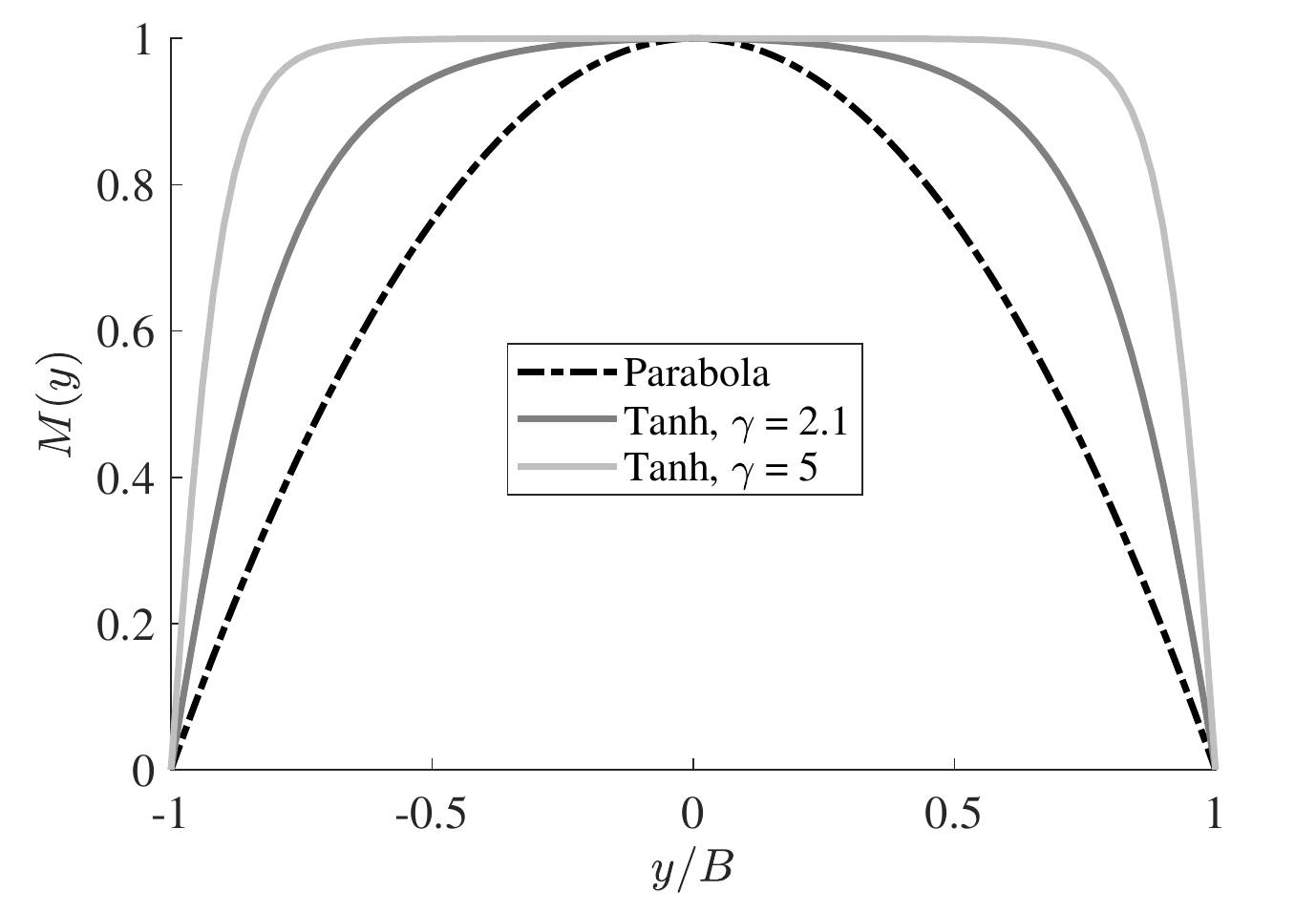}}
\caption{Illustration of the three different spanwise profiles $M(y) = M_p(y)$ (black dashed), $M_{2.1}(y)$ (dark gray solid) and $M_5(y)$ (light gray solid) used in the rest of the paper. The full base velocity is $U(y,z) = -\sin(\pi z) M(y)$.}\label{fig:BF} 
\end{figure}

Our analytical base flows $U(y,z)$ and $\mathcal{R}(z)$ were chosen to be both simple and realistic. 
For the velocity we considered: 
\begin{equation}
U(y,z) = -\sin(\pi z) M(y) \quad \quad \mbox{for}\ -1 \leq z \leq 1 \ \ \text{and} \ -B \leq y \leq B .
\label{eq:U_2D}
\end{equation}
The spanwise modulation  $M(y)$ can take one of the two following shapes: 
\begin{subeqnarray}
M_p(y) &=& 1-(y/B)^2 ,\slabel{eq:My_1}  \\
M_\gamma(y) &=& \frac{ \tanh \left[ \gamma \left( 1-(y/B)^2 \right) \right]  }{\tanh \gamma} . 
\slabel{eq:My_2}  
\end{subeqnarray}
Figure \ref{fig:BF}  shows the profiles $M_p,M_\gamma$ for the two different values of $\gamma$ used in this paper. 
Both profiles satisfy no-slip conditions at the walls. The  `Poiseuille' profile $M_p$ represents a steady, fully developed boundary layer extending throughout the entire y domain from one wall to the other. The `tanh' profile $M_\gamma$ has a parameter $\gamma$ whose increase above 1 generates increasingly flat profiles in the mid-plane $y=0$ and thin boundary layers at the walls. It models a flow whose spanwise boundary layer did not have sufficient time and/or length to develop fully. (The \textit{local} stability analysis requires the $Re$ number to be sufficiently large such that streamwise variations are on larger scales than the instability wavelength in order to remain relevant.) A fully-developed sine shape is used in $z$ in all cases, as this paper focuses on the effect of spanwise confinement. As a comparison, the experimental mean flow of \cite{Lefauve18} also has  roughly  a sine shape in $z$ (although slightly asymmetrically down-shifted), and our $M_\gamma$ in $y$ is an excellent approximation with a best fit obtained for $\gamma \approx 2.1$ . \newline

For the base density distribution, we considered the classical hyperbolic tangent $\mathcal{R}(z) = -\tanh \left[ (z-z_0)/\delta \right]$.  
This introduces two additional degrees of freedom: the density layer thickness ($\sim 2\delta$) and the asymmetry (or shift) parameter $z_0$.
The scaling of the sine profile Eq.($\ref{eq:U_2D}$) sets the shear layer thickness to $1$, leading to:
\begin{equation}
\mathcal{R}(z) = - \tanh\left[2 R (z-z_0) \right] \quad  \text{for} \ -1 \leq z \leq 1.
\end{equation} 
where we define $R= 1/(2 \delta)$ as the ratio of the shear layer thickness to the density layer thickness.

\subsubsection{Approach}

We are left with eight free parameters: $Re$; $Sc$; $Ri_b$; $\theta$; $z_0$; $R$; $M(y)$ and $B$. By choosing their values, and given a wavenumber $k \in \mathbb{R}^+$, all operators in Eq.(\ref{GenEig_2D}) are made fully explicit: the generalized eigenvalue problem can be solved numerically for both $\sigma(k) \in  \mathbb{C}$ and its associated eigenvector $[v,w,\rho,p]$. We express $\sigma(k) = \sigma_r(k) + i \sigma_i(k)$ where the subscripts $r$ and $i$ respectively denote the real part (growth rate) and the imaginary part. Therefore, here the phase velocity is $- \sigma_i(k)/k$ and the wave propagates in the the positive $x$-direction if it is positive. If $\sigma_r(k) >0 $  the wave is unstable  and grows as $\propto e^{\sigma_r t}$ until nonlinearities come into play (see \cite{Cudby21} for a treatment of these nonlinearities). \newline 

The eight free parameters are not all significant when focusing on the effect of spanwise confinement. The first key parameter for confinement is evidently the duct aspect ratio $B$. In addition, we  investigate the impact of viscous effects on both the base flow, by varying  $M(y)$ and $\gamma$, and also on the perturbation dynamics, by varying $Re$. As the base flow is not directly dependent on $Re$, both effects can be studied independently. Finally, in this paper, we also vary $Ri_b$,  the central  parameter for stratified shear instabilities. Therefore, in the remainder of the paper, the four remaining parameters will typically be set according to the experimental values of \cite{Lefauve18}: $(Sc,R,z_0,\theta) = (700,(1/0.047)/2,-0.22,5^{\circ})$.

\subsection{Summary of classical results ($1D$, unconfined)}

Since our study focuses on the effects of spanwise confinement, comparison with a classical (one-dimensional, unconfined, and spanwise-invariant) analysis is appropriate. In this problem,  which we refer to simply as the `$1D$ problem', all $y$-dependence is removed ($\p_y=0$) in Eq.(\ref{GenEig_2D}), and the base flow is  $U(z)=-\sin(\pi z)$. This leads to a simpler system, whose expression is given in Appendix \ref{1D_pb}. Note that by `$1D$' we do \textit{not} mean that the spanwise direction `does not exist', but that it is infinite and that no spanwise variations exist either for the base flow or the perturbations. This distinction should be kept in mind in the following.    \newline

As a foundation for our analysis of spanwise confinement, we now summarize the effects of the prominent parameters $(Re, Ri_b)$ on the `1D' stability properties. This short discussion results from supplementary `1D' computations, which for the sake of brevity are not illustrated. In addition, although fixed elsewhere, the effect of the parameter $z_0$ (quantifying the offset between the density and velocity base profiles) on the `1D' stability properties is briefly presented as well.
\begin{itemize}
     \item Effects of $Re$: The growth rates $\sigma_r$ increase monotonically and relatively uniformly with $Re$, until $Re \approx 1000$, where this effect tends to saturate.
    \item Effects of $Ri_b$: When the flow is unstratified, with $Ri_b=0$, the flow is only subject to a `pure' Kelvin-Helmholtz (denoted `$KH$') instability. As $Ri_b$ increases, the $KH$ mode is progressively weakened (i.e. $\sigma_r$ decreases), and eventually suppressed ($\sigma_r=0$), since, physically speaking, too much potential energy is required to allow the rolling up of the shear layer, and hence the density stratification. At $Ri_b \approx 0.125$ the flow then becomes subject to the inherently stratified Holmboe instability (denoted `$H$'), which is characterized at finite amplitude by propagating waves localized at the density interface ($z=z_0$), which generally counter-propagate. The growth rate of the Holmboe instability initially increases with $Ri_b$, up to a certain value, before $\sigma_r$ starts decreasing again. The most unstable wavenumber $k$ increases monotonically with $Ri_b$, since, in physical terms, longer waves require too much potential energy to allow instability. A physical mechanism based on wave interaction is now commonly proposed to explain the $KH$ and $H$ instabilities, as exhaustively reviewed in \cite{Carpenter13}, although the underlying arguments date back to G. I. Taylor's Adams Prize essay in 1915. Using a broken-line profile model for the shear layer and a localized density jump, the $KH$ instability can be interpreted as the interaction between the two counter-propagating vorticity waves localized at and `riding' their respective vorticity interfaces. In contrast, the $H$ instability appears as an interaction between one of the two vorticity waves, with one of the two gravity waves, that both `ride' the density interface. Discrimination is then made by noticing that, in order to interact, a wave pair must have \textit{intrinsic} phase speeds of opposed sign. This is shown in \citep{Carpenter13} as being equivalent to the Rayleigh theorem.  
    \item Effects of $z_0$: For $z_0=0$, the flow and the Holmboe instability are symmetric, in the sense that the distances between both vorticity waves and the density interface are equal. Both Holmboe waves are thus equally unstable and propagate with equal and opposite phase speeds. Mathematically, the corresponding eigenvalues are complex conjugates of each other. When $z_0 <0$, this symmetry is broken; the left-going wave ("$H_l$") becomes slower, of longer wavelength and more unstable, while the right-going wave ("$H_r$") becomes faster, of shorter wavelength and more weakly unstable (and \textit{vice versa}, there is a perfect symmetry in the case $z_0>0$ by swapping $H_r$ and $H_l$).
\end{itemize}

\section{Most unstable mode: preliminary observations}\label{sec2}
We now solve numerically the fully $2D$ generalized eigenvalue problem Eq.(\ref{GenEig_2D}), with confinement in both the crosswise and spanwise directions.  We choose a Poiseuille spanwise profile $M(y)=M_p(y)$, and parameters $(Re, Sc, R, z_0, \theta) = (440, 700,(1/0.047)/2, -0.22, 5^{\circ})$ (as in \cite{Lefauve18}) and three different aspect ratios $B=[1,3,5]$. The corresponding $1D$ (spanwise-invariant) problem is also solved for comparison (it can be viewed as the limit $B \rightarrow \infty$). It is not clear at this stage whether this $1D$  mode will be more unstable than $2D$ modes, since the duct walls create transverse shear in their vicinity (i.e. crosswise vorticity $\omega_z = \p_y U$), whose effect is, as yet, unknown. \newline 

\begin{figure}
\centering
  \begin{subfigure}[b]{0.48\linewidth}
    \includegraphics[trim={3.5cm 10cm 3.5cm 10cm},clip,width=\linewidth]{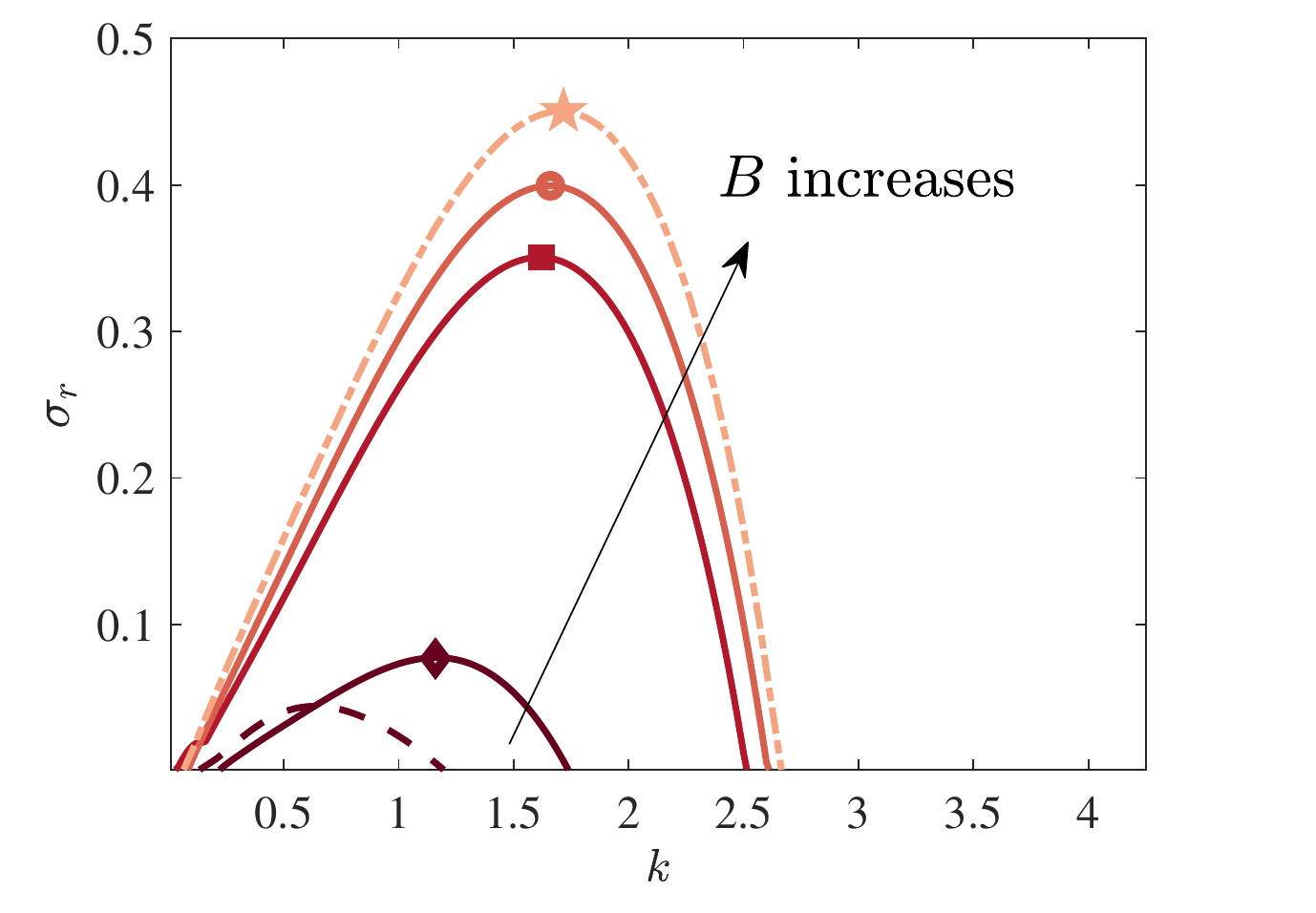}
     \caption{$Ri_b=0$ \label{fig:Ri0_DR_GR}}
  \end{subfigure}
  \begin{subfigure}[b]{0.48\linewidth}
    \includegraphics[trim={3.5cm 10cm 3.5cm 10cm},clip,width=\linewidth]{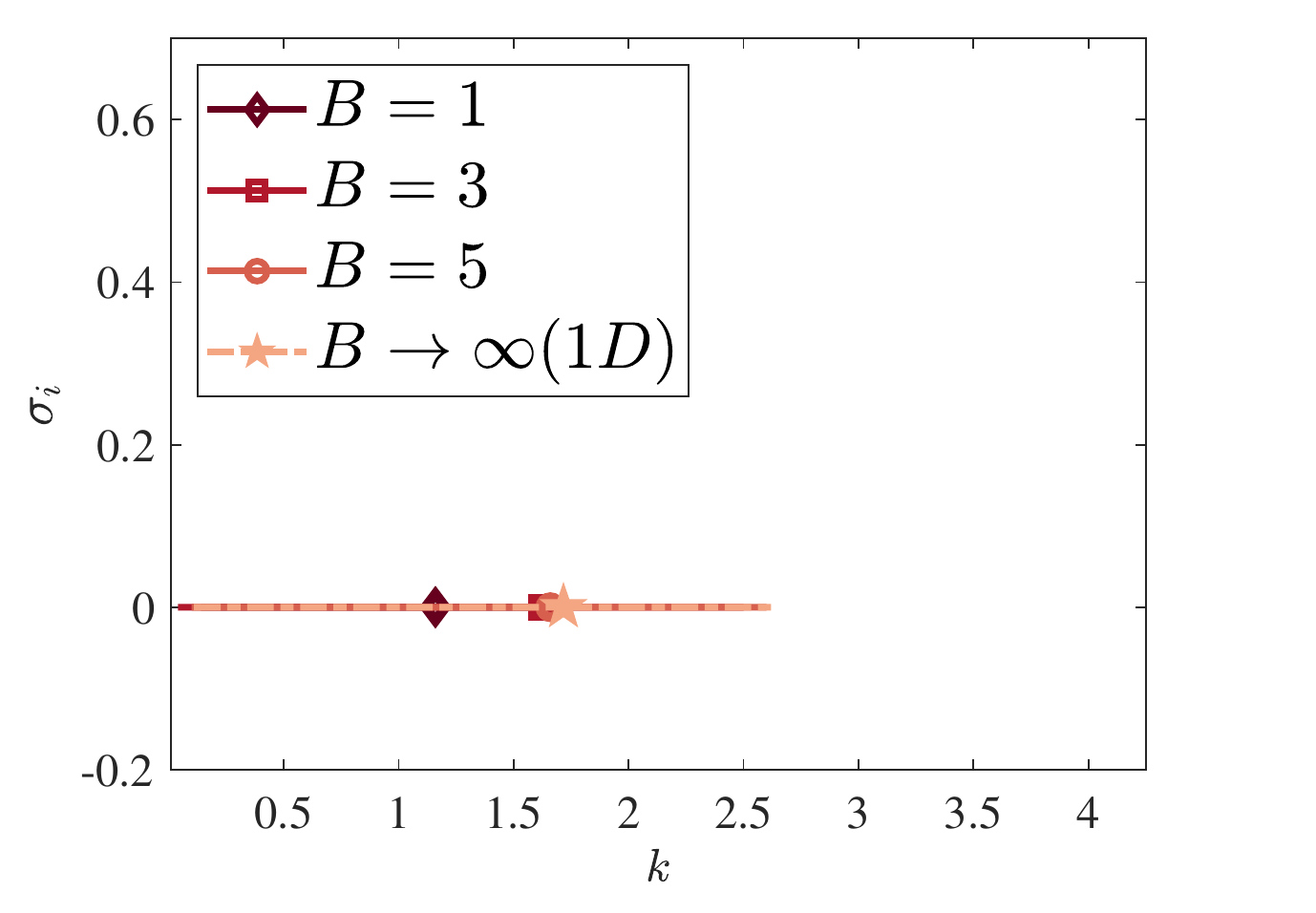}
    \caption{$Ri_b=0$ \label{fig:Ri0_DR_F}}
  \end{subfigure}
  \begin{subfigure}[b]{0.48\linewidth}
    \includegraphics[trim={3.5cm 10cm 3.5cm 10cm},clip,width=\linewidth]{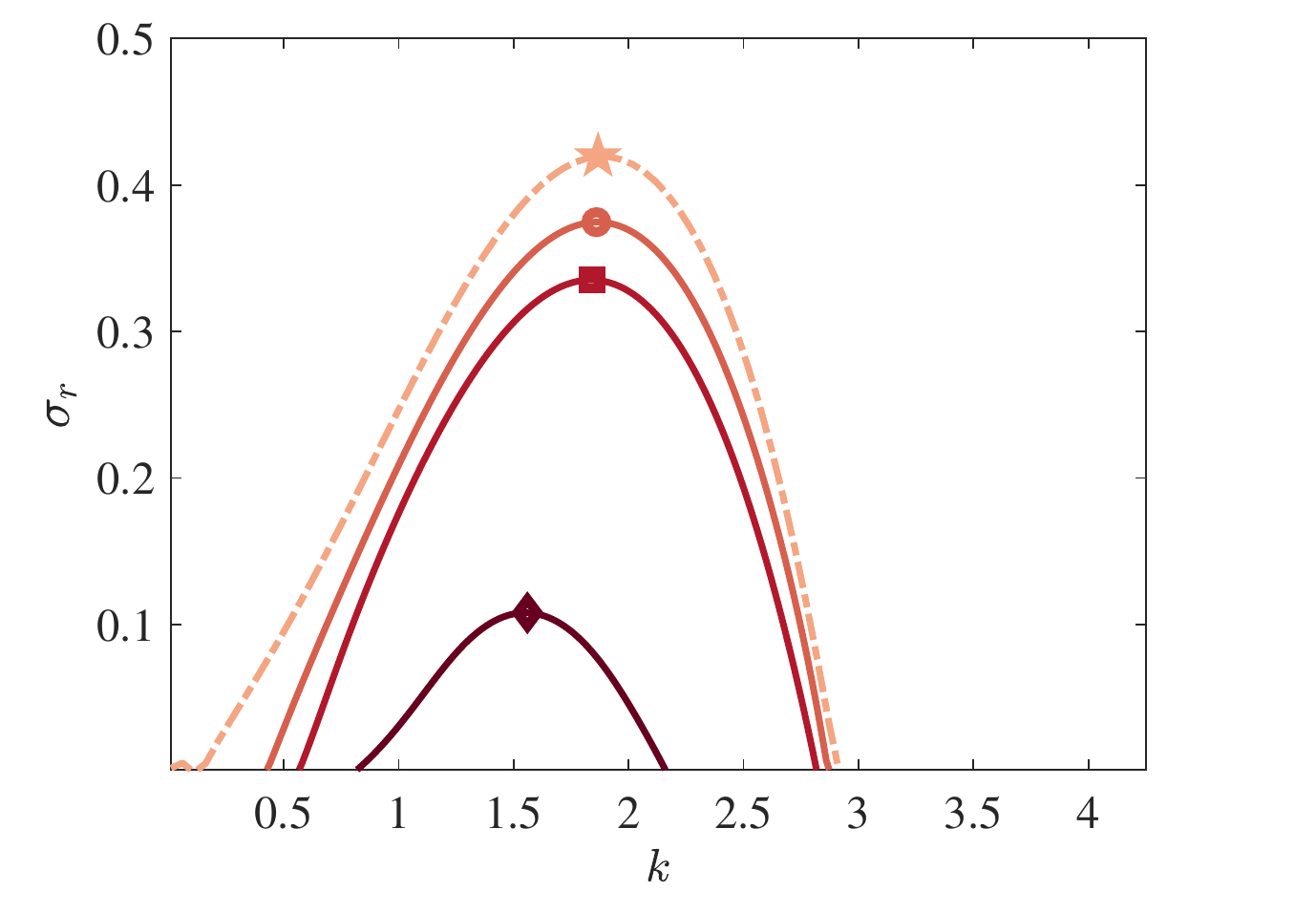}
    \caption{$Ri_b=0.25$ \label{fig:Ri0p25_DR_GR}}
  \end{subfigure}
  \begin{subfigure}[b]{0.48\linewidth}
    \includegraphics[trim={3.5cm 10cm 3.5cm 10cm},clip,width=\linewidth]{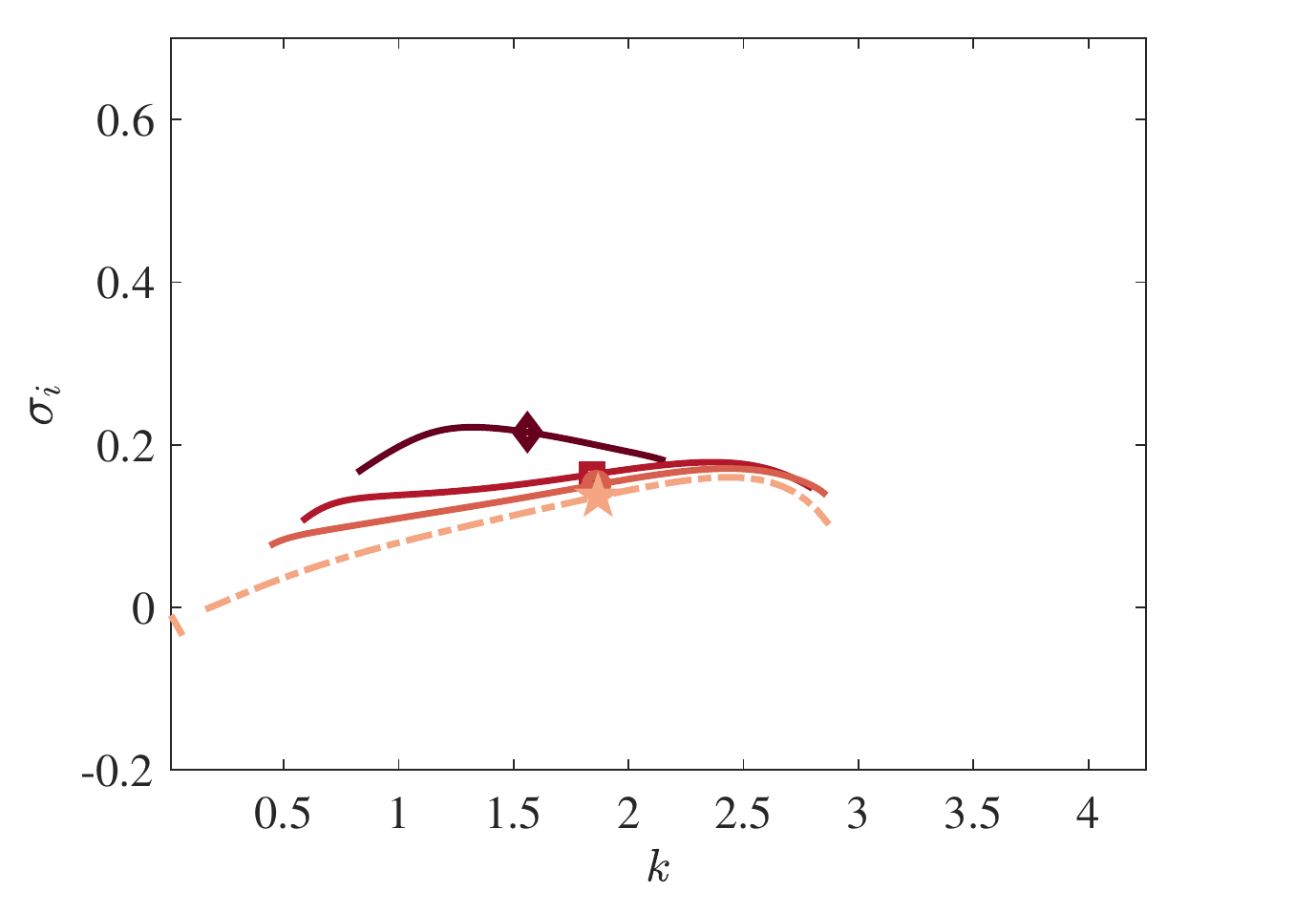}
    \caption{$Ri_b=0.25$ \label{fig:Ri0p25_DR_F} }
  \end{subfigure}
  \begin{subfigure}[b]{0.48\linewidth}
    \includegraphics[trim={3.5cm 10cm 3.5cm 10cm},clip,width=\linewidth]{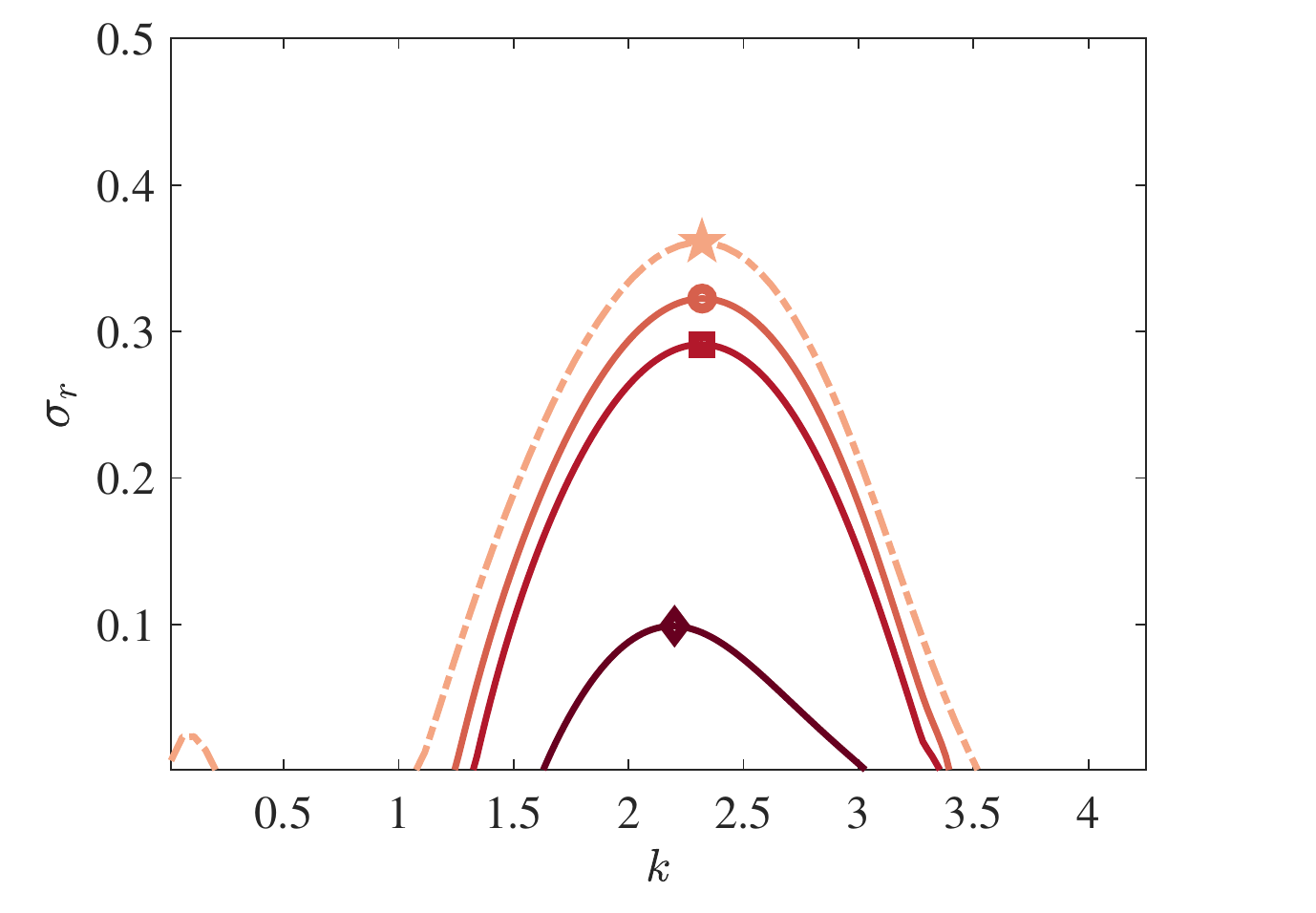}
    \caption{$Ri_b=1$ \label{fig:Ri1_DR_GR}}
  \end{subfigure}
  \begin{subfigure}[b]{0.48\linewidth}
    \includegraphics[trim={3.5cm 10cm 3.5cm 10cm},clip,width=\linewidth]{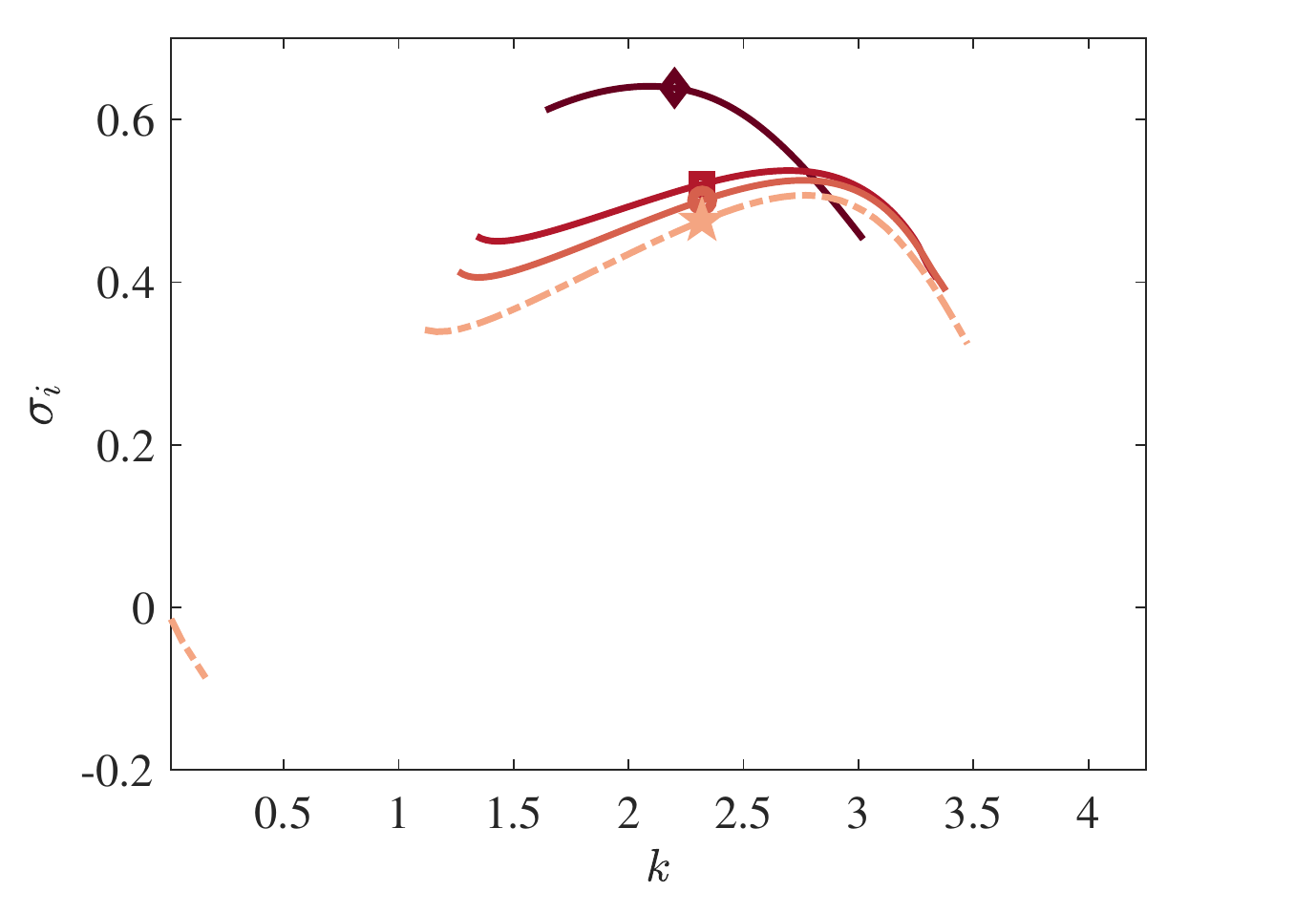}
    \caption{$Ri_b=1$ \label{fig:Ri1_DR_F}}
  \end{subfigure}
  \caption{Dispersion relations of the most unstable mode of the spectrum. The left column shows the growth rate $\sigma_r$, and the right column shows the frequency $\sigma_i$. We chose a Poiseuille spanwise profile $M_p$, four different aspect ratios $B$, and three different $Ri_b$ (rows). Solid and dashed lines stand for the $2D$ problem and lighter shade of grey corresponds to higher $B$ in the set $B=[1,3,5]$. Dash-dotted lines mark the $1D$ problem ($B \rightarrow \infty$). A marker, which is different for each $B$, indicates the maximum growth rate ($k=k_m$).}
  \label{fig:2D_DR}
\end{figure}

The dispersion relations of the \textit{most unstable} mode of the 2D spectrum are plotted with solid lines for $k=[0,4.5]$ in figure \ref{fig:2D_DR} for various
$B$ and $Ri_b$. For comparison, the 1D case (with $B \rightarrow \infty$) is plotted with a dot-dashed line.  Note that the most unstable mode (shown here with solid lines) has no reason to be the \textit{only} unstable mode for a given $k$. 
In  figure \ref{fig:Ri0_DR_GR}, the second most unstable unstratified mode is plotted with a dashed line for comparison for the case $B=1$. 
Furthermore, in the $1D$ stratified problem, we generally have two unstable $H$ modes, easily distinguishable by the sign of the phase speed. In the $2D$ problem, as $B$ increases, we also find an increasing number of unstable modes with similar phase speeds; however we defer this analysis to section \ref{sec:harmo} in order to first focus on the most unstable mode here.

%
%
%
%
%
%


\subsection{Kelvin-Helmholtz to Holmboe transition}

Figure \ref{fig:2D_DR} shows the following transitions between $KH$ and $H$ modes: 
\begin{itemize}
\item At $Ri_b=0$ (figures \ref{fig:Ri0_DR_GR}- \ref{fig:Ri0_DR_F}), the solid lines correspond to a $KH$ mode, while the dashed line (second most unstable mode) corresponds to a fully $3D$ mode,  which exists at $B=1$ and peaks at $k\approx 0.6$. To the authors' knowledge, this mode of instability is yet unclassified, and its characterization is deferred to section \ref{sec:Bmode}. Both the $KH$ mode and this new, fully $3D$ mode have zero phase speed.
\item At $Ri_b=0.25$ (figures \ref{fig:Ri0p25_DR_GR}-\ref{fig:Ri0p25_DR_F}), the dispersion relation corresponds to a $H_l$ mode (Holmboe traveling left). As shown in \cite{Carpenter10}, the frontier between $KH$ and $H$ modes at intermediate $Ri_b$ is blurred as soon as $z_0 \neq 0$, but we believe that $Ri_b = 0.25$ is sufficiently large for Holmboe modes to dominate unambiguously. As we choose a relatively strong negative asymmetry $z_0=-0.22$, the $H_r$ mode (traveling right) is absent (i.e. stable), just as in the $1D$ problem. 
\item At $Ri_b=1$ (figures \ref{fig:Ri1_DR_GR}-\ref{fig:Ri1_DR_F}), the previous $H_l$ mode remains. We also note the barely visible existence of a very weakly unstable $1D$ mode for $k\approx 0-0.2$ (also barely visible in figure \ref{fig:Ri0p25_DR_GR}), related to the nonzero tilting angle effect $\theta \neq 0$, the analysis of which is beyond the scope of this paper. 
\end{itemize}

We now discuss the effect of spanwise confinement on the dispersion relations. In the present temporal stability analysis ($k \in \mathbb{R}$ and $\sigma \in \mathbb{C}$), both  $KH$ and $H$ modes are stabilized by the spanwise confinement: values of $\sigma_r$ monotonically decrease below their $1D$ upper bounds for all wavenumbers $k$ as $B$ decreases. This stabilizing effect is more pronounced at small $B$. In this process, the most unstable wavenumber $k_m$ is approximately conserved, or perhaps very slightly decreased.  \newline

Note that the stabilizing effect of the sidewalls is in accordance with the results shown in \cite{Hocking78,Tatsumi90,Theofilis04}. These three studies establish the linear stability of an unstratified pressure-driven flow through a rectangular duct, namely a two-dimensional equivalent to the canonical plane Poiseuille flow. The analytical work \cite{Hocking78} considers the wide aspect ratio $B \gg 1$ limit, and performs a multiple asymptotic expansion based on the small parameter $B^{-1}$. In this asymptotic regime, a decrease in $B$ leads to an increase in the critical Reynolds number from the one-dimensional prediction $Re_{cr}=5772.22$. This conclusion is qualitatively supported by the numerically-oriented work \cite{Tatsumi90,Theofilis04}, that does not assume any particular regime for $B$. Physically, this stabilization was attributed in \cite{Hocking78} to a finite-Re effect, more precisely to the presence of spanwise boundary curvature in the base flow introduced by the sidewalls. Such physical interpretation remains to be verified in our case, as now attempted in section \ref{sec:stab}. \newline

\subsection{Convective to absolute instability transition}

Our temporal stability analysis has implications on the spatio-temporal properties of these flows. With increasing confinement (decreasing $B$), longer waves (traveling left) are sped up, evidenced by the  $\sigma_i$ curves being shifted up at low $k$ in figure \ref{fig:Ri0_DR_F}, \ref{fig:Ri0p25_DR_F} and \ref{fig:Ri1_DR_F}. The shortest waves, on the other hand, are slowed down, presumably because they encounter more significant viscous effects. An interesting consequence of this selective speed up and slow down is that the group velocity of the unstable wave-packet centroid, defined as:
    \begin{equation*}
    V_m = -\frac{\mathrm{d} \sigma_i}{\mathrm{d} k}(k_m),
    \end{equation*}
flips its sign as $B$ is reduced from $\infty$ to $0$. This means that there exists a value of $B$ such that the wave-packet centroid is static.  \newline

For example at $Ri_b = 0.25$ (figure \ref{fig:Ri0p25_DR_F}), the slope at the marker $\mathrm{d}\sigma_i/\mathrm{d}k(k_m)$ vanishes somewhere between $B=3$ and $B=1$. In the meantime, it is clear in figure \ref{fig:Ri0p25_DR_GR} that the flow remains unstable. This implies that, for the chosen set of parameters, confinement effectively renders the flow absolutely unstable, since an unstable wave-packet centered in $V_m = 0$ necessarily corresponds to an absolute instability. For $Ri_b = 0.25$, where the derivative is always of a small amplitude, the $H_l$ wave is very likely to be absolute for all the $B$ shown. However, for $Ri_b=1$, it may be convective for $B=5$, whereas it is certainly absolute for some $B\in [1,3]$. In this latter case, we conclude that spanwise confinement would \textit{destabilize} the flow in a spatio-temporal sense. A rigorous saddle-point \cite{briggs64,bers75,Huerre90,carriere99,Juniper06} or impulse response \cite{Brancher97,Delbende98,del98,Gallaire03} approach would be interesting in future work as discussed further in section \ref{sec:conc}.

\subsection{Stabilization by side walls and base flows \label{sec:stab}}

We previously attributed the stabilizing influence of confinement to a finite-$Re$ effect. To investigate this further, we analyse separately the effects of viscosity on (i) the base flow  and (ii) the perturbations. For (i) we keep $Re=440$ but switch the spanwise profile from $M_p(y)$ to $M_{2.1}(y)$ and $M_5(y)$ (decreasing the boundary layer thickness, see figure \ref{fig:BF}). For (ii) we keep $M(y)=M_p(y)$ but double $Re$ to $Re=880$. \newline

To quantify the (temporally) stabilizing effect of spanwise confinement, we define $E_r$ as the relative difference between the $1D$ most unstable eigenvalue $\sigma_m^{1D}=\sigma^{1D}(k_m^{1D})$ and the $2D$ eigenvalue evaluated at the same ($1D$ most unstable) wavenumber $\sigma (k_m^{1D})$: 
\begin{equation*}
E_r(B) = \frac{\left \| \sigma_m^{1D} - \sigma(k^{1D}_m ; B)  \right \|}{\left \| \sigma_m^{1D} \right \|}.
\end{equation*} 
Choosing $\sigma (k_m^{1D}) $ instead of $\sigma_m$  allows us to avoid solving the $2D$ eigenvalue problem for several $k$ at each $B$ (in order to find $k_m$). In addition, we are interested in the limit $B\rightarrow \infty$ where it is clear in figures \ref{fig:Ri0_DR_GR}, \ref{fig:Ri0p25_DR_GR} and \ref{fig:Ri1_DR_GR} that $k_m^{1D}$ becomes an excellent approximation of the $2D$ most unstable wavenumber $k_m$ for $B \gg 1$ (already for $B=3$, and even more so at the strong stratification $Ri_b =1$). \newline

\begin{figure}   
\centering
  \begin{subfigure}[b]{0.49\linewidth}
    \includegraphics[trim={3.5cm 10cm 4.2cm 10.2cm},clip,width=1\linewidth]{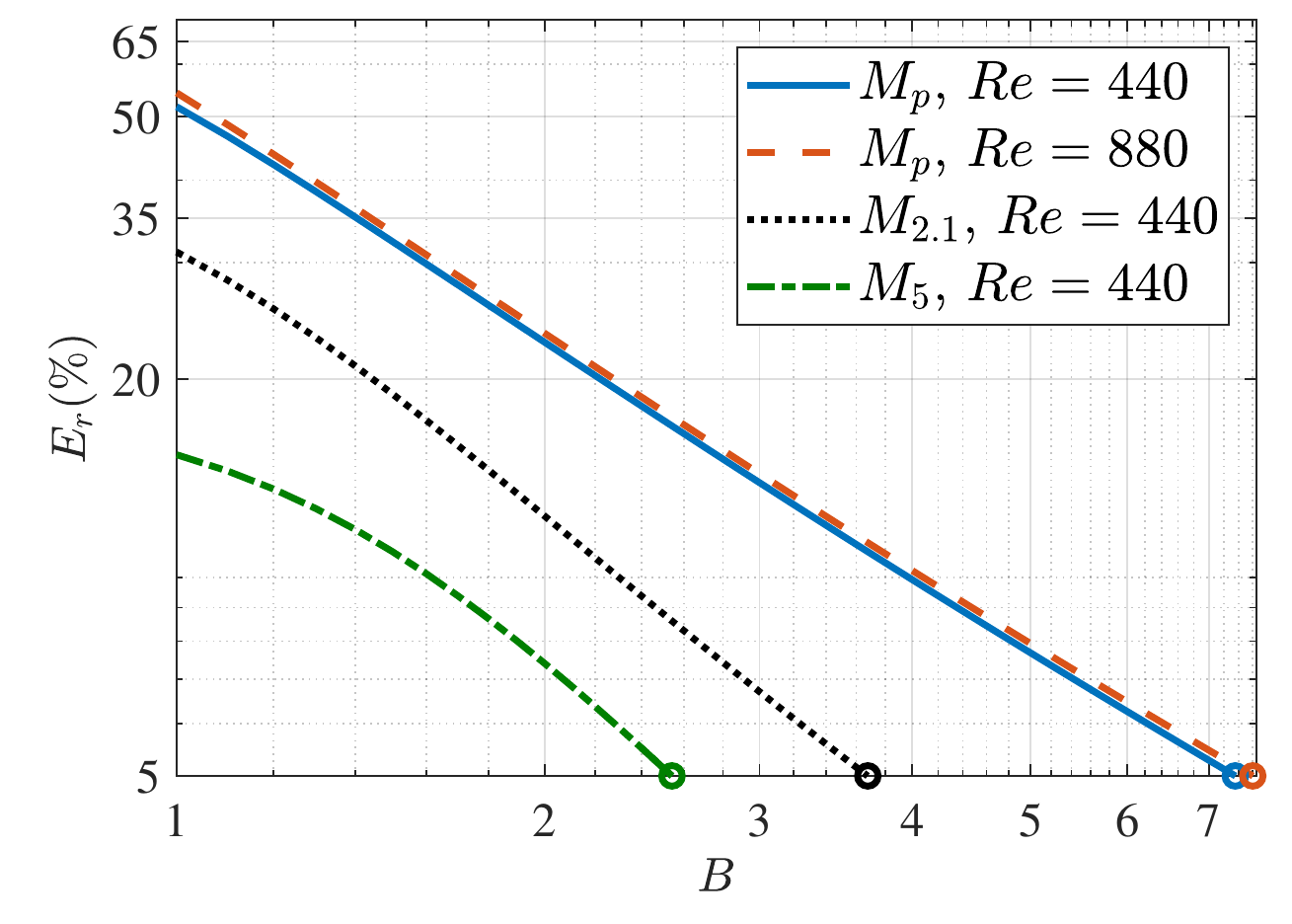}
  \caption{\label{fig:ErB_Ri1}}
  \end{subfigure}
  \hfill
  \begin{subfigure}[b]{0.49\linewidth}
     \includegraphics[trim={3.5cm 10cm 4cm 10cm},clip,width=1\linewidth]{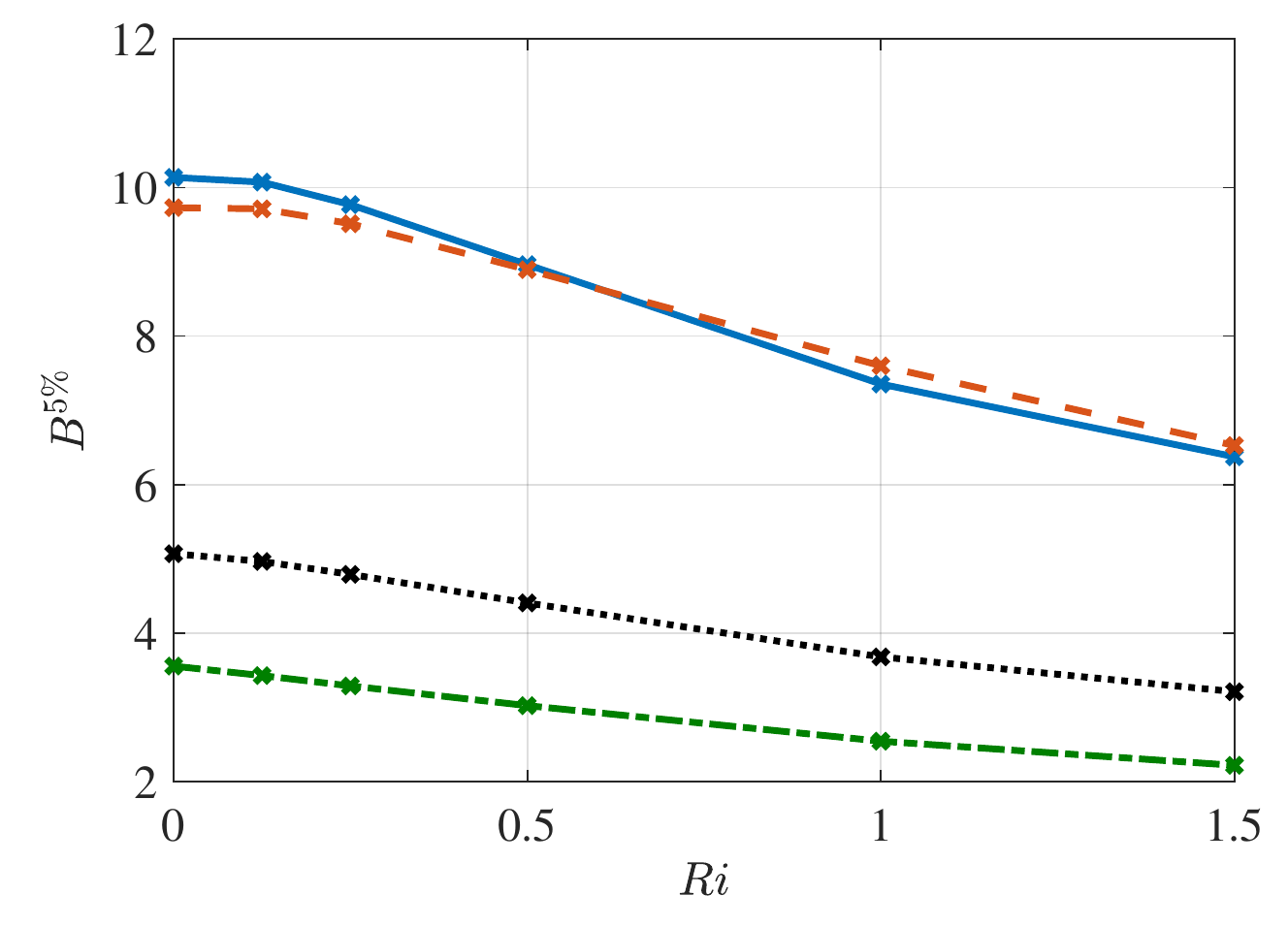}
  \caption{\label{fig:B5pc}}
  \end{subfigure}
    \caption{Stabilizing effects of the aspect ratio $B$, spanwise profile $M(y)$, and $Re$. (a) Relative error between the $1D$ and $2D$ most unstable growth rate evaluated at the most unstable $1D$ wavenumber $k_m^{1D} \approx 1.88$. Three different spanwise profiles and two different $Re$ are chosen, all for $Ri_b=1$.  A circle symbol is placed at the threshold aspect ratio $B^{5 \, \%}$ where the error is $E_r = 5 \% $; (b) Variation of this threshold aspect ratio $B^{5 \, \%}$ with $Ri_b$.}
     \label{fig:ErB}
\end{figure}

In figure \ref{fig:ErB_Ri1} we therefore plot $E_r(B)$ (in percentage terms)  in the four cases considered, setting $Ri_b=1$. Only  differences greater than $E_r\ge5 \%$ are shown since lower values required computationally prohibitive $B$ values.
First, we see that, for a given $Re=440$, difference curves for the tanh profiles $M_\gamma$ are significantly lower than for  the Poiseuille profile $M_p$, and even more so for larger values of $\gamma$. In other words, base flows with thinner boundary layers yield growth rates $\sigma_r$ that are less affected (in the specific sense of being less stabilized) by side wall confinement. Interpreting the $1D$ problem as an unbounded and constant $M(y)=1$, it appears natural indeed to expect convergence of $\sigma$ for a base flow that resembles $M(y)=1$ over the longest $y$ interval, which is here $M_5(y)$ (followed by $M_{2.1}$, and finally $M_p$).  
Interestingly, this also suggests that in these stratified shear instabilities, the spanwise boundary layers have a pure stabilizing effect. In broad terms, the boundary layer structure simply decreases the amount of kinetic energy available from the base flow to feed the instability, without introducing a viscous instability mechanism (e.g. Tollmien-Schlichting waves), at least at the values of $Re$ considered therein. \newline

Second,  we observe in figure \ref{fig:ErB_Ri1} that  both dash-dotted lines for $Re=440$ and $880$, almost collapse on each other. That indicates that, when viscous diffusion affects the perturbation \textit{alone}, its impact on the convergence towards the $1D$ problem is very weak.  We conclude that if viscous effects have indeed generally a strong damping impact on  stability properties, it is most significantly through their indirect effect on the base flow rather than through their direct effect on the perturbations dynamics alone. \newline 

We are now interested to know if the previous observations remain true for other values of $Ri_b$. Therefore in figure \ref{fig:B5pc},
we focus on the evolution with $Ri_b$ of the threshold aspect ratio $B^{5 \%}$, for which the error is $E_r(B^{5 \%}) = 5 \%$ (highlighted by a circle on the $x$ axis of figure \ref{fig:ErB_Ri1}, where we set $Ri_b=1$). We see that our conclusions for $Ri_b=1$ remain valid for other values of $Ri_b\in [0, 1.5]$. 
We further learn from figure \ref{fig:B5pc} that this threshold aspect ratio is reduced with increasing stratification: the pure $KH$ mode at $Ri_b=0$ is the most affected by the stabilizing influence of boundary layers, whereas the $H$ mode at high $Ri_b$ appears to easily match its $1D$ counterpart, i.e. it is least affected by confinement. This might be linked to the fact that higher $Ri_b$ are linked to shorter wavelengths which naturally tend to be less affected by the relatively more distant walls. In the next section we will see that confinement heavily affects Holmboe modes in more subtle ways, through the creation of spanwise harmonics.
\section{Spanwise harmonic Holmboe modes \label{sec:harmo}}

\subsection{Eigenvalue spectra and oblique mode analysis}

\begin{figure}
\centering
  \begin{subfigure}[b]{0.48\linewidth}
    \includegraphics[trim={3.5cm 10cm 3.5cm 10cm},clip,width=\linewidth]{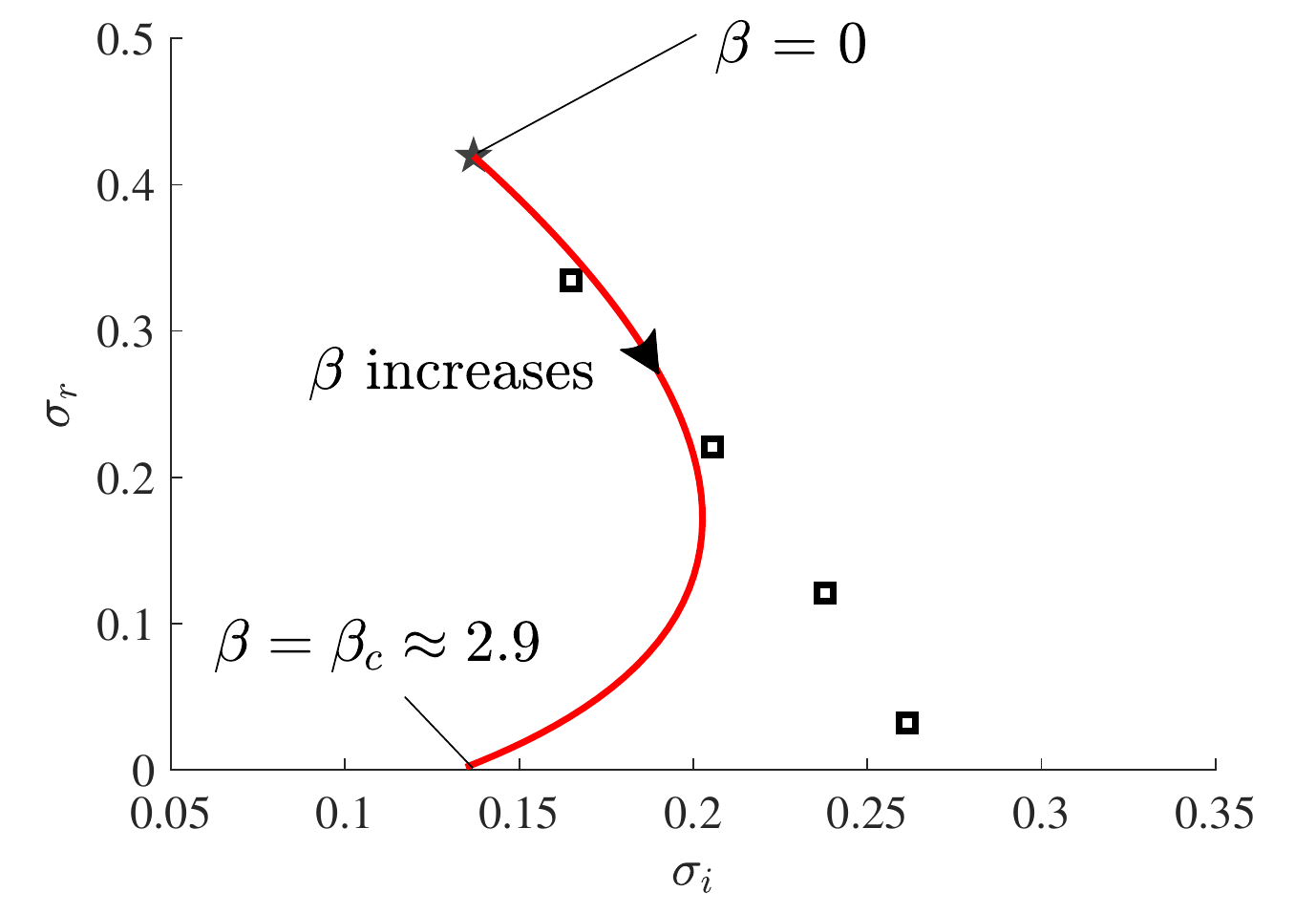}
    \caption{ $B=3$, Poiseuille \label{fig:S_B3} }
  \end{subfigure}
  \begin{subfigure}[b]{0.48\linewidth}
    \includegraphics[trim={3.5cm 10cm 3.5cm 10cm},clip,width=\linewidth]{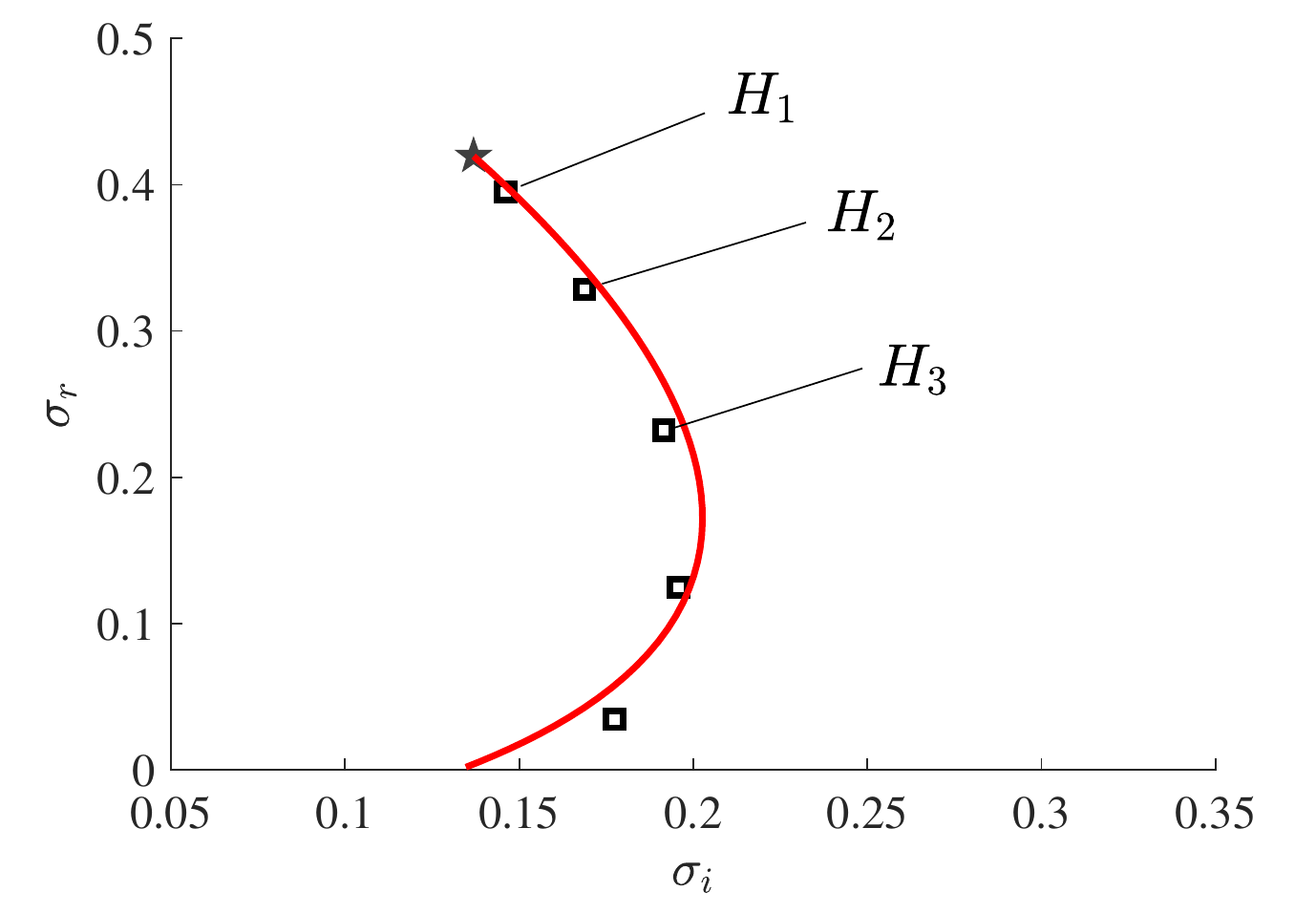}
    \caption{ $B=3$, tanh with $\gamma=5$ \label{fig:S_B3_tanh} }
  \end{subfigure}
  \begin{subfigure}[b]{0.48\linewidth}
    \includegraphics[trim={3.5cm 10cm 3.5cm 10cm},clip,width=\linewidth]{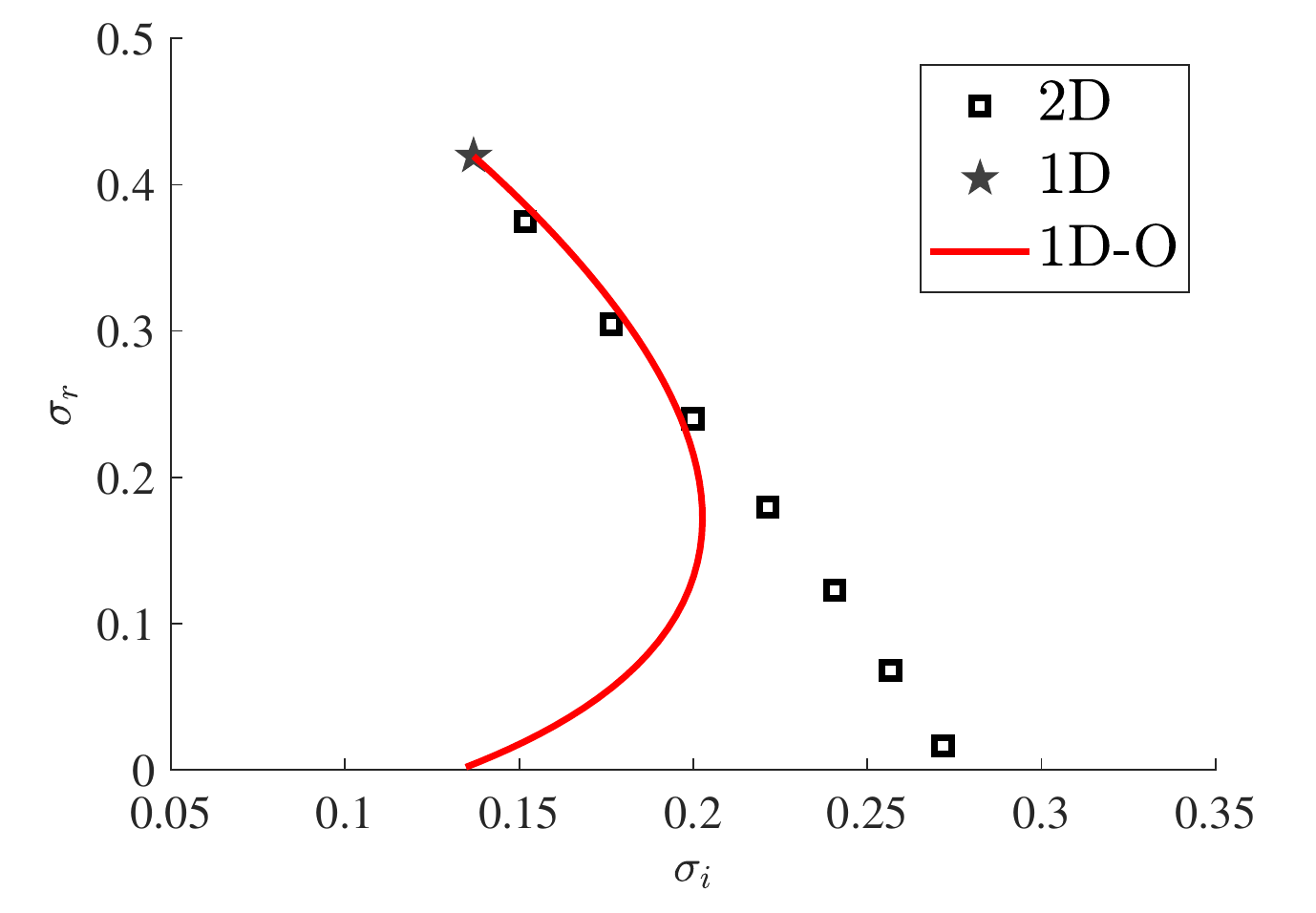}
    \caption{ $B=5$, Poiseuille \label{fig:S_B5} }
  \end{subfigure}
  \begin{subfigure}[b]{0.48\linewidth}
    \includegraphics[trim={3.5cm 10cm 3.5cm 10cm},clip,width=\linewidth]{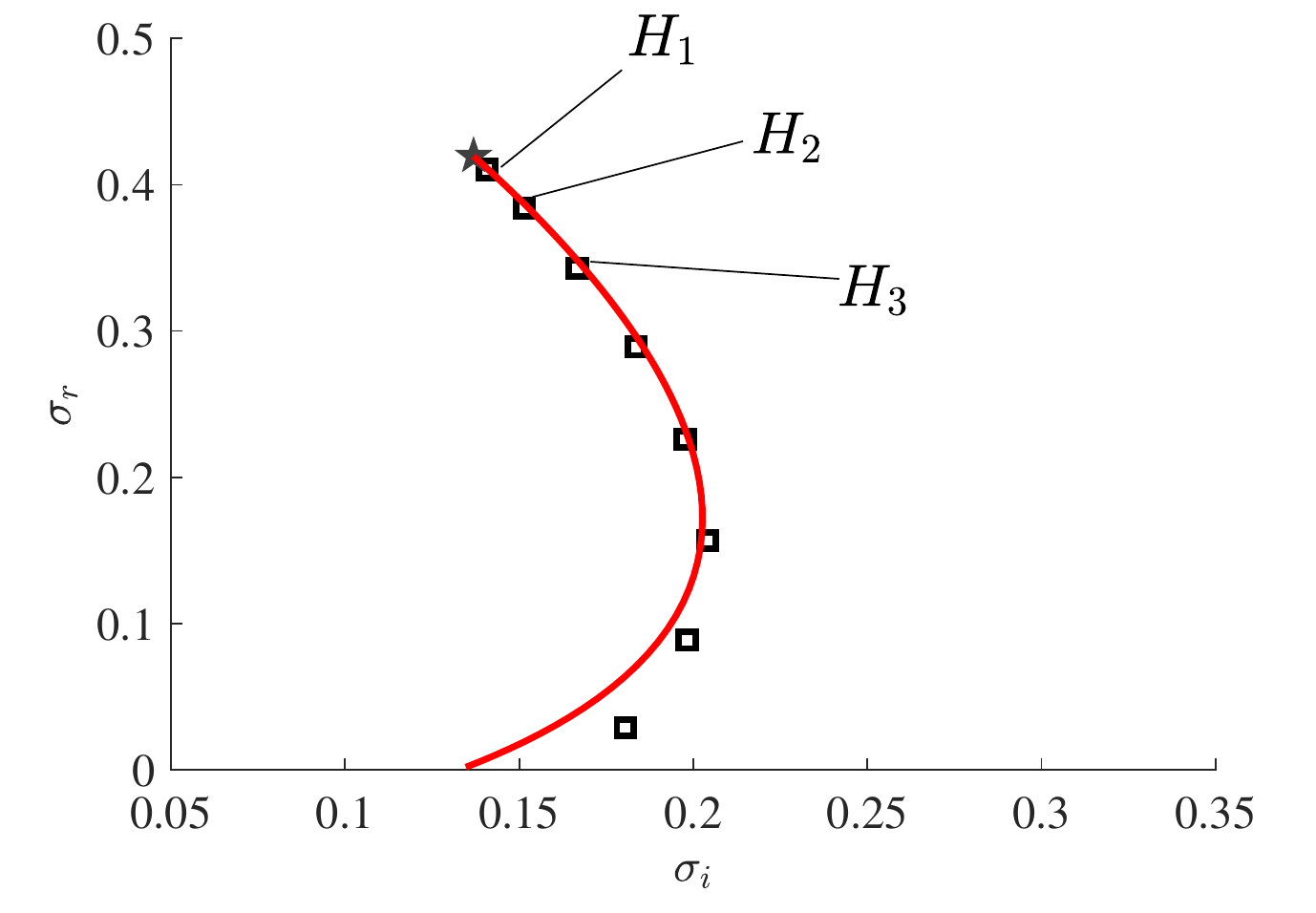}
    \caption{ $B=5$, tanh with $\gamma=5$ \label{fig:S_B5_tanh} }
  \end{subfigure}
  \caption{Unstable part of the spectra for $Re=440$, $Ri_b = 0.25$, $k=k_m^{1D} = 1.88$. We show two different aspect ratio flows $B=3, \, 5$  and two different  $M=M_p, \, M_5$. The black star symbol marks the $1D$ eigenvalue, the black squares are the $2D$ eigenvalues, and the red line is the $1D-O$ dispersion relation as a function of the spanwise wavenumber $\beta$. The latter is the same for the four panels as it does not consider the spanwise variations of the base flow. As a consequence of the strongly negative $z_0$, note that all eigenvalues are in the $\sigma_i >0$ half plane.} \label{fig:Spec}
\end{figure}

Heretofore, given a wavenumber $k$, only the most unstable mode of the eigenvalue spectrum was represented in figure \ref{fig:2D_DR} (and used in figure \ref{fig:ErB}). Whereas this mode is unique in the $1D$ problem as a consequence of the strong asymmetry (quantified by $z_0$) that stabilizes the opposite Holmboe wave, it has no reason to be in the $2D$ problem. In fact, there exists a fairly important number of unstable modes besides the most unstable one; they are now shown in figure \ref{fig:Spec}. In the left column, we show the unstable part of the spectrum (black squares) for a Poiseuille spanwise base flow  ($Re=440, Ri_b=0.25$ and $B=3,5$ corresponding to figures~\ref{fig:Ri0p25_DR_GR}-\ref{fig:Ri0p25_DR_F}). In the  right column, we replace $M_p(y)$ by $M_5(y)$. The $1D$ eigenvalue is systematically plotted (with black stars) for comparison.
To understand the existence of these multiple unstable modes, we consider and superimpose (shown with a red curve) the dispersion relation of oblique modes, found by taking a spanwise independent flow $U(z) = -\sin( \pi z)$ (in other words $M(y)=1$), and  expanding any perturbations field $f$, as:
\begin{equation} \label{eq-oblique-mode}
\hat{f}(y,z) = \overline{f}(z)\exp{(i\beta y)}, 
\end{equation} 
where $\beta \in \mathbb{R}$ is the spanwise wavenumber. A single mode (corresponding to a choice of $k,\beta$) is a propagating wave whose front is perpendicular to $(k\mathbf{\hat{x}} + \beta \mathbf{\hat{y}})$. It is clear that the eigenvalues $\sigma_r + i \sigma_i$ are now also functions of $\beta$. As $\beta \in \mathbb{R}$, all wavelengths are allowed for the perturbations, so that this description implicitly assumes that the spanwise direction is unbounded. Its also requires the coefficients of the equations to be independent on $y$, so we must have, as introduced earlier, a spanwise-invariant base flow, i.e $M(y)=1$. This problem is therefore similar to the $1D$ one, except that perturbations are allowed to vary in $y$. In the rest of the paper, this problem will thus be referred to as the $1D-O$ problem (`$O$' for oblique). It is not \textit{a priori} clear whether the $\beta=0$ (i.e. the $1D$) eigenvalue is the most unstable one among all possible $\beta$. A stratified generalization of Squire's theorem (\citep{Squire33}), presented in \citep{Smyth90}, states that a $1D-O$ mode  ($\beta \neq 0$) has a smaller growth rate $\sigma_r$ than a corresponding $1D$ mode ($\beta=0$) having lower $Re$ and a larger $Ri_b$. However, since $H$ modes generally have $\sigma_r$ increasing with $Ri_b$, this theorem is inconclusive  in the present context and does not preclude the dominance of $1D-O$ modes over $1D$ modes (for more details see \cite{Lefauve18} \S~2.3.3). \newline

Let us now observe the $2D$, $1D-O$ and $1D$ unstable spectra, focusing first on the left column of figure \ref{fig:Spec} ($M = M_p$). At $B=3$ (panel a), four $2D$ eigenvalues are unstable. As the aspect ratio is increased to $B=5$ (panel c), these eigenvalues approach the $1D$ value and three new distinct unstable $2D$ eigenvalues appear by crossing the real axis, bringing the total to seven eigenvalues `originating' from the $1D$ eigenvalue. As a matter of fact, the $1D$ eigenvalue is always more unstable than $2D$ ones. Oblique modes (red line) of course match the $1D$ eigenvalue for $\beta=0$; as $\beta$ increases, they draw a `comma' shape in the complex plane down to a cut-off $\beta = \beta_c$ where they cross the real axis and become stable. Note that  the $1D-O$ eigenvalues, just as the $2D$ eigenvalues, never become more unstable than the $1D$ eigenvalue: the instability does not take `advantage' of the wave-front rotation, as it does for instance in Tollmien-Schlichting instability \citep{Schmid12}. Oblique modes, and their associated spanwise curvature, simply undergo stronger viscous damping as $\beta$ is increased. Indeed, the Laplacian term $Re^{-1}(-k^2+\partial_{zz}+\partial_{yy})$ becomes $Re^{-1}(-k^2+\partial_{zz}-\beta^2)$ under the oblique mode expansion (see appendix \ref{1Dobl_pb}) ; thus, larger $\beta$ give more weight to this diffusive term. Consequently, $\beta_c$ is primarily determined by the value of $Re$; for the parameters of figure \ref{fig:Spec}, we obtain $\beta_c \approx 2.9$. 
\newline

Focusing now on the right column of figure \ref{fig:Spec} ($M = M_5$), we observe exactly the same process as $B$ is increased, except that $2D$ unstable modes become more numerous (five in panel b \emph{vs} four in panel a, and eight in panel d \emph{vs} seven in panel c), and more tightly packed around the $1D$ one. More interestingly however, the $2D$ eigenvalues fit much better the oblique mode dispersion relation, in particular at $B=5$ (panel d), where the black squares appear to be nothing else than a discretized version of the red curve, with only a slight error for the more stable modes. \newline

Note that an emergence from the $1D$ eigenvalue of an increasing number of inherently $2D$ eigenvalues as the aspect ratio is varied  was already observed in \cite{Theofilis04} for  Poiseuille flow. However, to the authors' knowledge, a quantitative comparison with the oblique dispersion relation is novel.

\subsection{Spatial structures and symmetries}

\begin{figure}
\centering
    \includegraphics[trim={0cm 1.5cm 0cm 8cm},clip,width=0.85\linewidth]{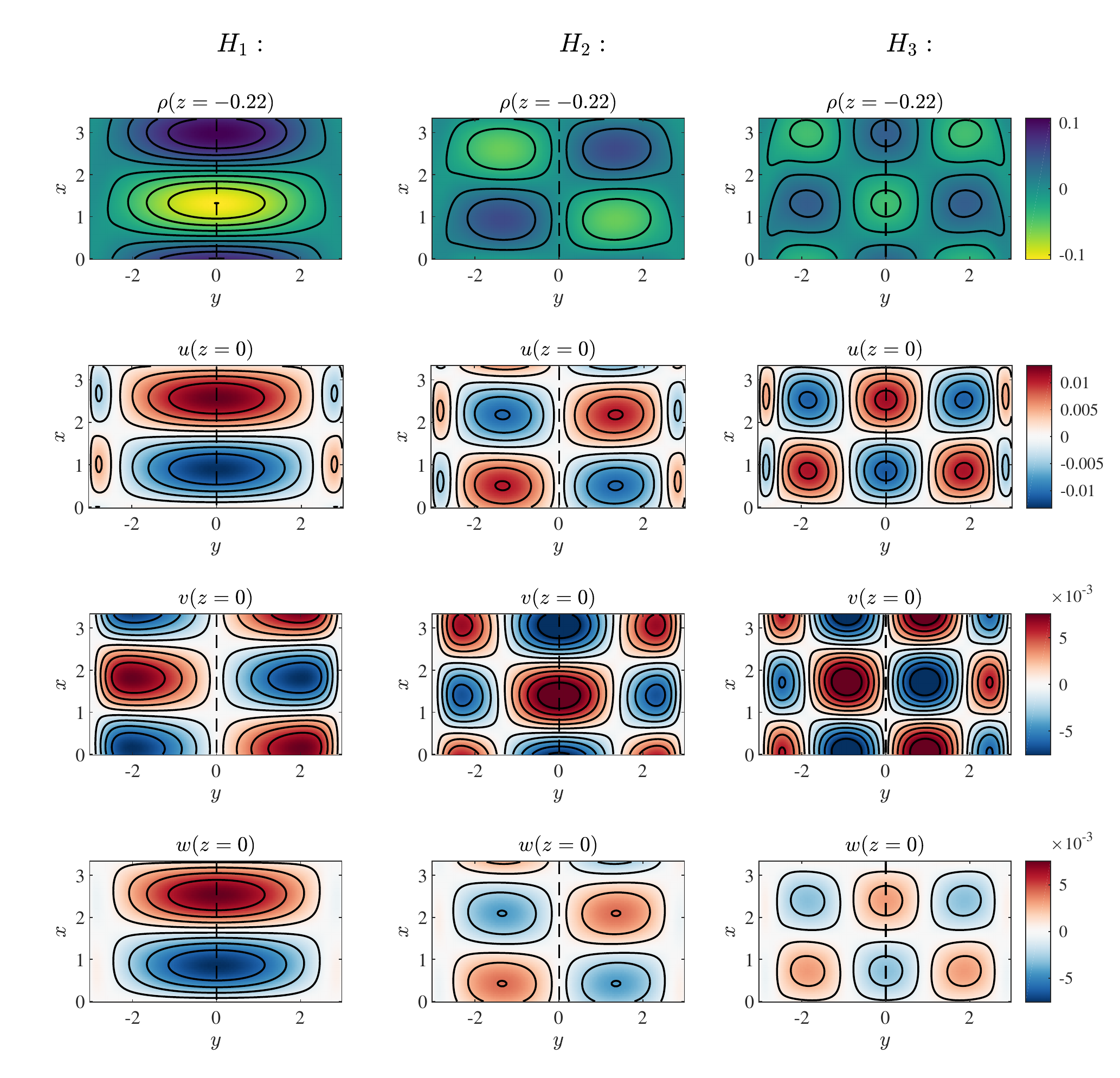}
  \caption{Sliced views of $\rho,u,v,w)$ in the $x-y$ plane of the three  most unstable $2D$ eigenmodes of figure~\ref{fig:S_B3_tanh} ($H_1$, $H_2$ and $H_3$, from left to right, ordered by decreasing growth rates). The spanwise profile is $M(y)=M_5(y)$, and the parameters are $Re = 440$, $Ri_b=0.25$, $B=3$, $k=k_m^{1D} = 1.88$. }
\label{fig:Harmomics}
\end{figure}

To understand this phenomenon better, we show in figure \ref{fig:Harmomics} the spatial structure of the three most unstable eigenmodes of figure \ref{fig:S_B3_tanh}, labeled $H_1$, $H_2$, and $H_3$, by decreasing growth rates. Each column shows an $x-y$ slice of the eigenmode (in the  plane $z=0$ for velocities, and $z=z_0$ for the density). Although insufficient for a phenomenological understanding of the instability, this visualization allows us to compare the spanwise structure of eigenmodes with each other. \newline

\begin{figure}
\centering
    \includegraphics[trim={0cm 1.5cm 0cm 16cm},clip,width=1\linewidth]{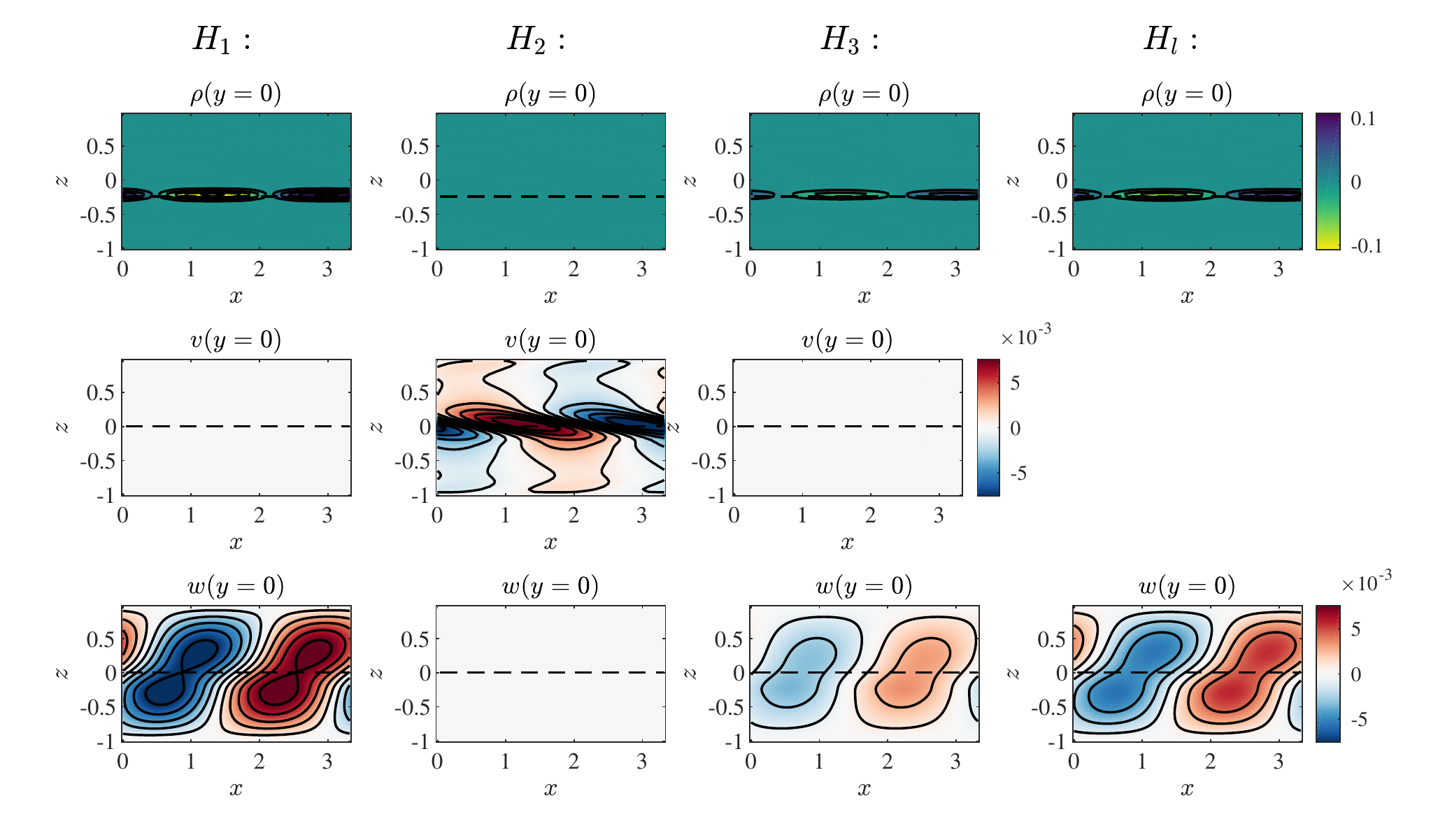}
  \caption{Sliced view on the $x-z$ plane of the three $2D$ most unstable eigenmodes for $k=k_m^{1D} = 1.88 $. The $1D$ most unstable mode is added at the fourth column.  Modes from left to right between the first and the third column are $H_1$, $H_2$ and $H_3$, as ordered in decreasing order of growth rates. $M(y)=M_5(y)$, $Re = 440$, $Ri_b=0.25$ and  $B=3$}
\label{fig:HarmomicsYS}
\end{figure}

Two different types of \textit{spanwise} symmetries are revealed. The first symmetry, that we call $S_1$, can be characterized as (even,odd,even,even) for $(u,v,w,\rho)$ respectively. Namely, the field $u(y,z)$ is even around the axis $y=0$, such that $u(-y,z) = u(y,z)$ ; meanwhile, the field $v(y,z)$ is odd around the axis $y=0$, such that $v(-y,z) = -v(y,z)$, etc... The second symmetry, $S_2$ is its opposite: (odd,even,odd,odd). In both cases, $v$ has a different symmetry from all other perturbation components. It can be checked that these two symmetries are indeed allowed by Eq.(\ref{GenEig_2D}) as long as both $U$ and $ \mathcal{R}$ are even in $y$. We complete figure \ref{fig:Harmomics} with figure \ref{fig:HarmomicsYS}, showing $x-z$ slices of the same three modes in the orthogonal plane $y=0$. In the last column, we add the $1D$ mode $H_l$ for comparison.  
Simultaneous observation of figure \ref{fig:Harmomics} and \ref{fig:HarmomicsYS} allow us to draw the following conclusions. 
\begin{enumerate}
    \item The most unstable 2D mode, $H_1$, has symmetry $S_1$. It appears to be a simple `$2D$ extension' of the $1D$ mode for $\rho$, $u$ and $w$. Indeed, its  $x-z$ structure at $y=0$  is extremely similar to the $1D$ structure, and we remember that symmetry $S_1$ guarantees that $v=0$ everywhere on this plane (as in the 1D case). 
     In addition, no particular spanwise variation is observed for $\rho$, $u$ and $w$ (figure \ref{fig:Harmomics}) other than those required to match the boundary conditions at the walls. This mode should be essentially seen as a two-dimensional version of $H_l$, and is identical to the confined Holmboe instability of \cite{Lefauve18}.
    \item The second most unstable eigenmode $H_2$  has symmetry $S_2$, of which no $1D$ equivalent  exists (compare the second and the fourth columns of figure \ref{fig:HarmomicsYS}). The density interface is now also wavy in the spanwise direction. Structurally, this $H_2$ mode should be seen as a `harmonic in $y$' of the first mode $H_1$. Indeed $u,w,\rho$ have a spanwise `periodicity' of $2B$ (the quotes indicate that a true periodicity clearly cannot be satisfied because of the boundary conditions), in comparison to $4B$ in $H_1$; furthermore $v$ is now $(4B/3)$-periodic compared to being $(2B)$-periodic in $H_1$.
    \item The third most unstable eigenmode $H_3$ goes back to symmetry $S_1$, and is one step further in the harmonic range. The spanwise periodicity is now $(4B/3)$ for  $u,w,\rho$ and $B$ for $v$.
\end{enumerate}
This emerging logic of alternating symmetries, as a consequence of higher-order harmonics, extends to higher modes : the fourth most unstable mode $H_4$ has symmetry $S_2$, the fifth has symmetry $S_1$, etc. Overall, if we rank modes by decreasing order of growth rate, then the $i\text{th}$ mode, $H_i$, is $S_1$ if $i$ odd and $S_2$ if $i$ even. Spanwise `wavelengths' (the quotes again  draw attention to  the fact that the shape is not purely sinusoidal) are summarized as:
\begin{subeqnarray} \label{eq:lvw}
\lambda_i^{\rho} = \lambda_i^u  = \lambda_i^w &=& \frac{4B}{i}  \qquad \text{for} \ i=1,2,3,... \slabel{eq:lw} \\
\lambda_i^v &=& \frac{4B}{i+1} \ \quad \text{for} \ i=1,2,3,... \slabel{eq:lv}
\end{subeqnarray}
It is important to note that this mode structural-ordering is conserved as we change $B$. Because of this spanwise-`periodic' shape of $2D$ modes, we now understand the (imperfect) alignment of corresponding eigenvalues on the oblique modes dispersion relation observed in figure \ref{fig:S_B3_tanh} (and other panels). This may appear surprising since $2D$ modes propagate purely along $x$, whereas oblique modes make an angle $\beta$ with the background flow. However, it is shown in appendix \ref{1Dobl_pb} that modes that are periodic standing waves in $y$ and that travel purely along the streamwise $x$-direction satisfy the $1D-O$ dispersion relation (thanks to the symmetry of the system). These modes are:
\begin{equation*}
\begin{cases}
\hat{h}(y,z) = \breve{h}(z)\cos(\beta y)  &  \\ 
\hat{v}(y,z) = \breve{v}(z)\sin(\beta y) &  
\end{cases}   \text{with symmetry} \  S_1,  \ \  \text{and} \ \  
\begin{cases}
\hat{h}(y,z) = \breve{h}(z)\sin(\beta y)  &  \\ 
\hat{v}(y,z) = \breve{v}(z)\cos(\beta y) & 
\end{cases}   \text{with symmetry}  \ S_2 ,
\end{equation*}
where `$h$' denotes  $u$, $w$, $\rho$ or $p$ (the hat and the breve are simply added to distinguish these particular mode shapes from the one arising directly from (\ref{GenEig_2D})). Of course, the $2D$ eigenvalue spectrum necessarily remains a  \textit{discrete} version of this  $1D-O$ dispersion relation  (continuous in $\beta$) since only a few `wavelengths' satisfy the boundary conditions due to quantization. \newline

\begin{figure}
\centering
  \begin{subfigure}[b]{0.48\linewidth}
    \includegraphics[trim={3.5cm 10cm 3.5cm 10cm},clip,width=\linewidth]{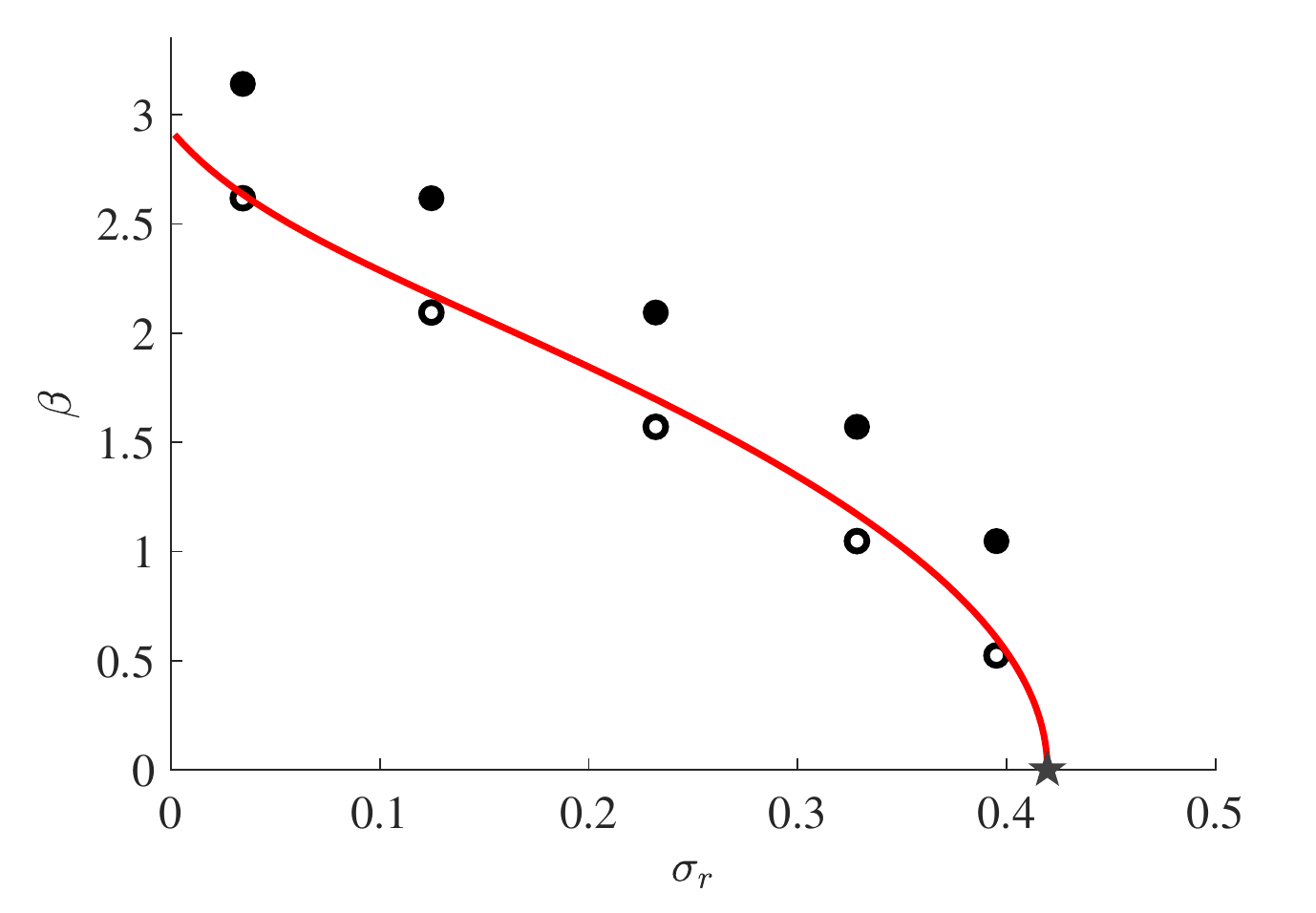}
     \caption{$B=3$ \label{fig:S_B3_tanh_REAL} }
  \end{subfigure}
  \begin{subfigure}[b]{0.48\linewidth}
    \includegraphics[trim={3.5cm 10cm 3.5cm 10cm},clip,width=\linewidth]{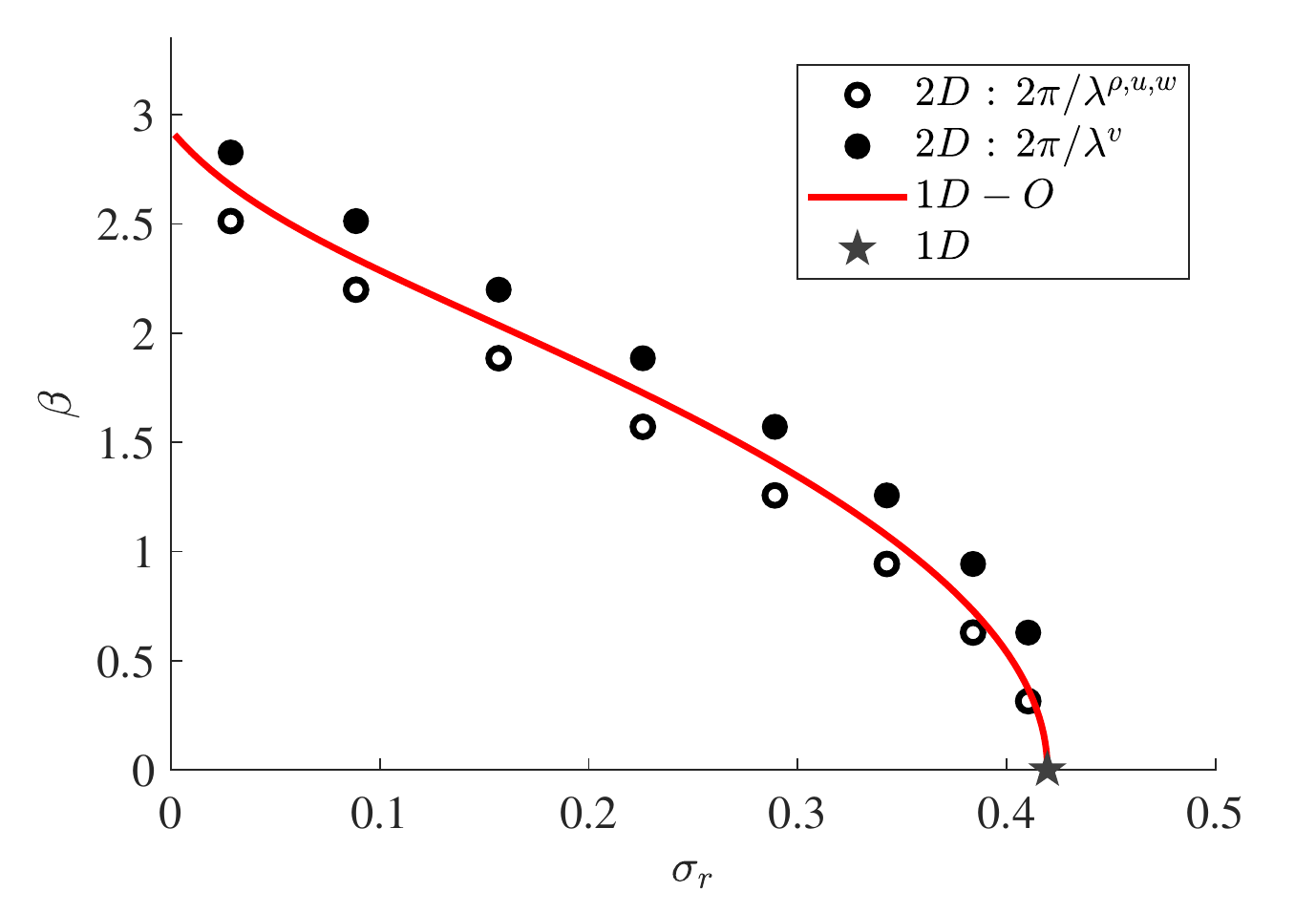}
    \caption{$B=5$ \label{fig:S_B5_tanh_REAL} }
  \end{subfigure}
    \caption{Growth rates of oblique modes as a function of $\beta$ (red line), and of $2D$ modes at $k=k_m^{1D}=1.88$ as a function of \textit{both} $\lambda_i^{\rho,u,w} $ (empty circles) and $\lambda_i^{v}$ (full circles). Parameters are $M(y) = M_5(y)$, $Re = 440$ and $Ri_b=0.25$. The $1D-O$ dispersion relation in $\beta$ predicts the discrete $2D$ spanwise harmonics increasingly well as $B$ is increased and $M_5(y) \rightarrow 1$ over most of the domain.}
    \label{fig:SpecREALIMAG}
\end{figure}

To illustrate this point, we plot in figure \ref{fig:SpecREALIMAG} the wavelengths $\lambda_i^{\rho,u,w}$ and $\lambda_i^{v}$ of the $2D$ modes together with the one predicted by the $1D-O$ dispersion relation, as a function of the growth rate ($\beta$ \emph{vs} $\sigma_r$ plot). For $B=3$ (panel a) the agreement between  $2D$ and $1D-O$ `wavelengths' is very good, simply because $2D$ modes possess structures akin to $1D-O$ standing waves in $y$. The slight discrepancy between $2D$ and $1D-O$ growth rate, previously observed in figure \ref{fig:S_B3_tanh}, may now be explained as follows. In the $1D-O$ problem, only one wavelength $\lambda = 2\pi/\beta$ is predicted for a given $\sigma_r$, and it is equal for all fields $u,v,w,\rho$. By contrast, in the $2D$ problem, we already reported that $\lambda_i^{v} <\lambda_i^{\rho,u,w}$. This mismatch is an interesting consequence of the system symmetry, to which is added the no slip boundary conditions at $\pm B$.  To be more specific, let us first focus on $H_1$ (the most unstable mode in figure \ref{fig:Harmomics}). The corresponding $1D-O$ mode would predict a $v$ that is phase-shifted by $\pi/2$ in $y$ with respect to all the others fields; thus, $|v|$ would be $0$ in the middle of the duct and maximum at $y=\pm B$, violating the no-slip boundary conditions. Therefore, the $2D$ mode adapts by decreasing its wavelength by the least possible amount so that the right and left lobes of $v$ cancel at the walls. A similar phenomenon happens in all the other unstable modes. The $2D$ growth rate is then determined by a compromise between both wavelengths. Consequently, reversing the paradigm and fixing a value of $\sigma_r$ in figure \ref{fig:SpecREALIMAG}, we can say that the $1D-O$ wavelength is above  $\lambda_i^{v}$ and below $\lambda_i^{\rho,u,w}$. 

Last but not least, oblique modes require a $y$-invariant base flow $M(y)=1$ whereas the tanh $2D$ base flows have boundary layers near the wall. This additional source of discrepancy between $2D$ and $1D-O$ modes is clearly illustrated by comparing figure \ref{fig:S_B5} ($B=3$) to figure \ref{fig:S_B5_tanh} ($B=5$): the alignment of $2D$ eigenvalues on the $1D-O$ dispersion relation is clearly improved, and becomes very good, by reducing the relative boundary layer thickness. 
Moreover, increasing $B$ improves the alignment and makes the $1D-O$ model increasingly accurate. Indeed, the difference between the $u$,$w$,$\rho$ and $v$ wavelengths tends to $0$ as $B$ increases, since:  
\begin{equation}
     \frac{1}{\lambda_i^{v}} - \frac{1}{\lambda_i^{\rho,u,w}}  = \frac{1}{4B} .
\end{equation}
Since this difference is caused by the nonmatching of boundary conditions of $v$ at the walls, it is logically attenuated as they are moved away. We indeed see that full and empty circles are closer to each other in figure \ref{fig:S_B5_tanh} (compared with figure \ref{fig:S_B3_tanh}),  and in figure \ref{fig:S_B5_tanh_REAL} (compared with figure \ref{fig:S_B3_tanh_REAL}). \newline

\subsection{Importance of spanwise harmonics at weak confinement ($B \rightarrow \infty$)}

We now address the question of higher harmonics becoming unstable, and/or even more unstable, as $B$ is increased. Considering Eq.(\ref{eq:lvw}), the answer comes naturally: the `wavelength' of the $i\text{th}$ mode increases with $B$. Physically, the mode is stretching out as the walls are moved away. Consequently, the `wavenumber' $\beta_i^{2D} = 2 \pi / \lambda_i^{v}$ or $2 \pi / \lambda_i^{\rho,u,w}$ decreases, but we saw with the $1D-O$ analysis (figure \ref{fig:S_B3_tanh_REAL} or \ref{fig:S_B5_tanh_REAL} ) that lower $\beta$ correspond to higher $\sigma_r$ since such modes experience less viscous damping. In the limit $B \rightarrow \infty$, we expect the number of unstable modes to be infinite since $\Delta \beta_i^{2D} \propto B^{-1}\rightarrow 0$: we can have an infinite number of $i$ before reaching the viscous cut-off $\beta_c$. In other words, in this limit, the discrete set of $2D$ unstable eigenvalues becomes a continuous spectrum, as one should expect from a Fourier \textit{transform} in an infinite domain, as opposed to a Fourier \textit{series} in a bounded domain. Moreover, in this limit, the $2D$ unstable spectrum is expected to \textit{become} the $1D-O$ one if we choose $M(y)=1$.  \newline

The above comments, although generally expected and relatively unsurprising, may have interesting implications for linear stability analyses at large aspect ratios  $B\rightarrow \infty$. Namely, provided the base flow is almost $y$ invariant far from the boundaries and has no $v$ velocity, we conclude the following : 
\begin{enumerate}
\item The set of $1D-O$ eigenvalues for  $\beta = \frac{2 \pi i}{4B}$ with $i \in \mathbb{Z}$ gives an excellent prediction of the $2D$ spectrum. 

\item The $2D$ spectrum becomes increasingly denser; $2D$ unstable eigenvalues are numerous, and the most unstable ones are very close, even though they correspond to modes with different spatial structures. As a practical consequence, if a given spanwise eigenmode is preferentially excited (for whatever reason) one may observe a pattern that is completely different from that predicted by the $1D$ analysis.  
\end{enumerate}

\section{`Twisted Kelvin-Helmholtz' mode}\label{sec:Bmode}

\subsection{Dominance of a new mode $KH_T$ at low $k$ and $Rib\approx 0$}

\begin{figure}
\centering
  \begin{subfigure}[b]{0.7\linewidth}
    \includegraphics[trim={5.2cm 16.5cm 5.2cm 3.75cm},clip,width=0.95\linewidth]{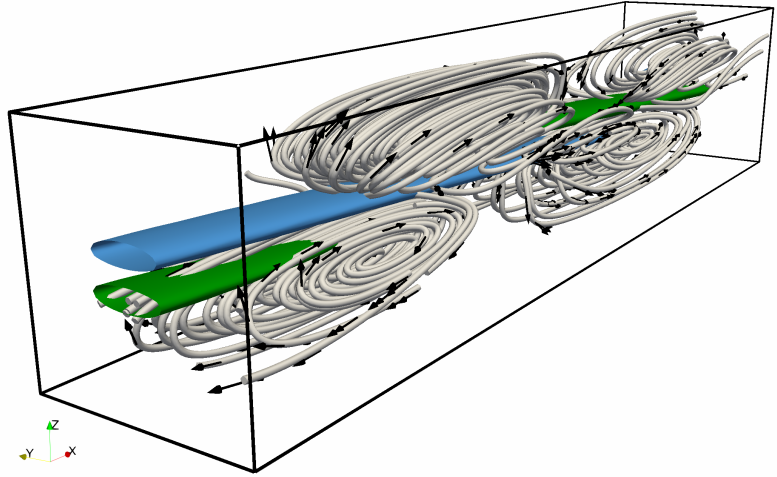}
     \caption{ $KH_1$, where blue and green surfaces are isocontours of equal and opposite values of $\omega_y$   \label{fig:3Dview_KH}}
  \end{subfigure}
  \begin{subfigure}[b]{0.7\linewidth}
    \includegraphics[trim={5.2cm 16.5cm 5.2cm 3.7cm},clip,width=0.95\linewidth]{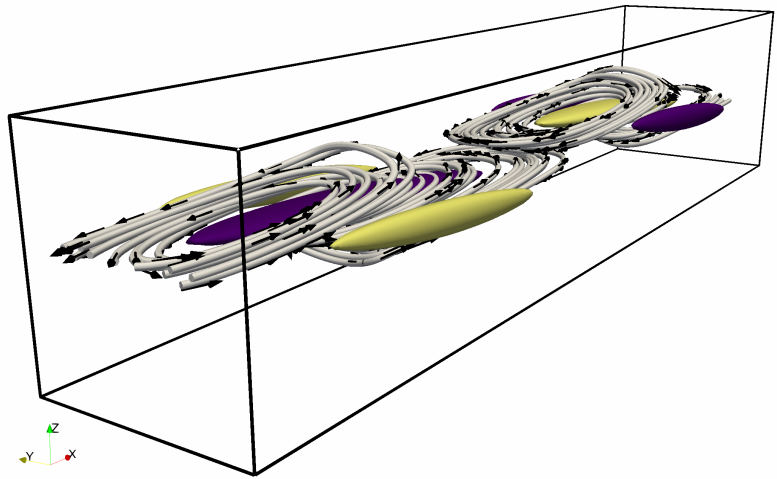}
    \caption{ $KH_T$, where yellow and magenta surfaces are isocontours of equal and opposite values of $\omega_z$ \label{fig:3Dview_KH_H2} }
  \end{subfigure}
    \caption{Visualizations of the $KH_1$ and $KH_T$ eigenmodes for $B=1$ and $k = 0.64$ where $\sigma_r(k = 0.64) \approx 0.0440$ ; ($M=M_p(y), Re=440, Ri_b=0$), highlighting their different spatial structure. }
 \label{fig:3Dview_KHKH2}
\end{figure}

\begin{figure}
\centering
    \includegraphics[trim={0.5cm 6cm 0cm 9.2cm},clip,width=1.05\linewidth]{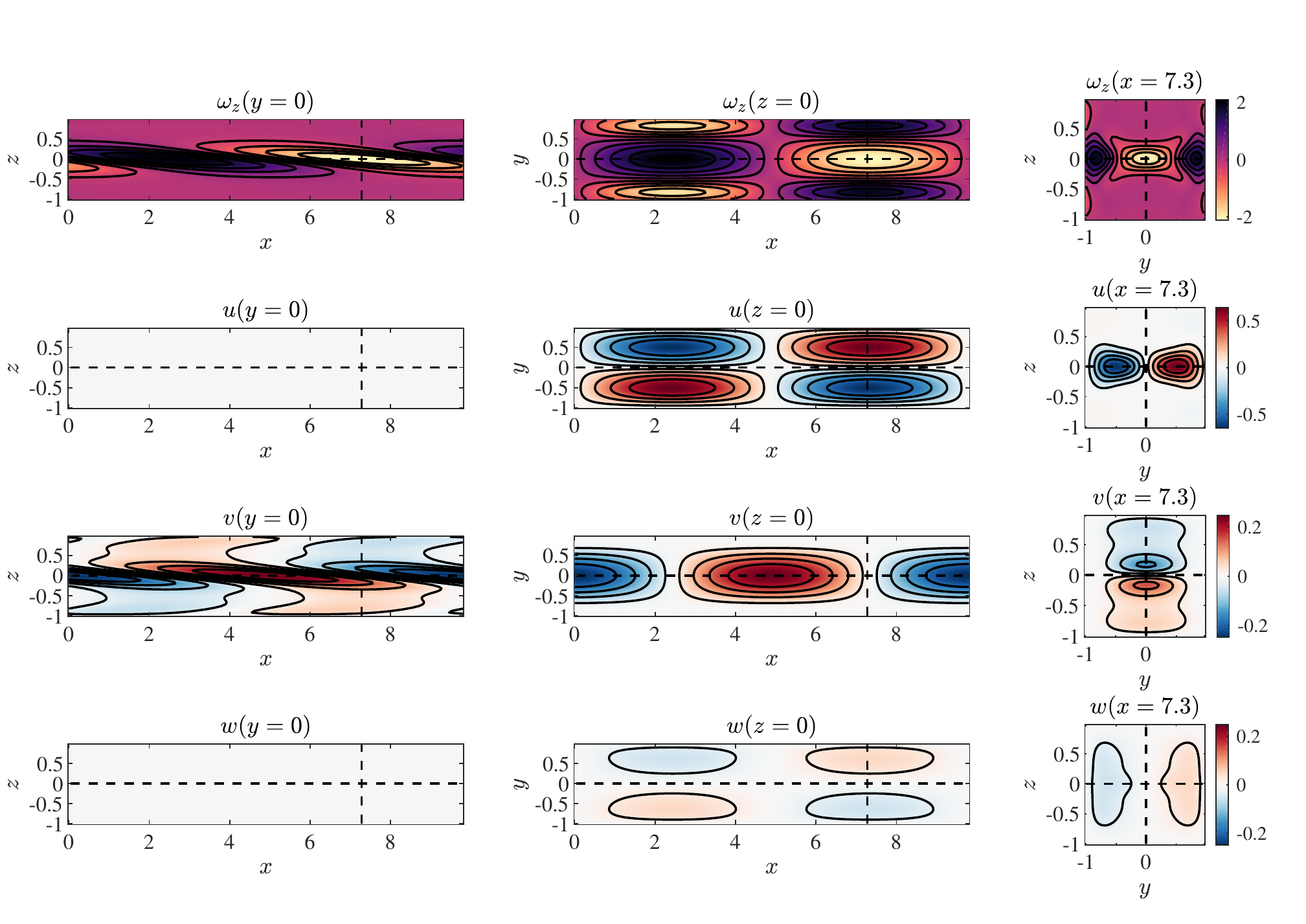}
  \caption{Some slice views of the $KH_T$ mode shown in figure \ref{fig:3Dview_KH_H2}. Dashed lines show the locations of the planes in the other columns. 
  }
  \label{fig:Slice_KH_H2}
\end{figure}

For particular combinations of $Ri_b$, $k$ and $B$ (everything else being held fixed), it may occur that one of the previously described simple '$2D$-extension' of a $1D$ mode is not the dominant one. The dashed lines in figure \ref{fig:Ri0_DR_GR} ($Ri_b=0$) show that it can indeed happen. This phenomenon actually  appears inherently unstratified, and quickly disappears as $Ri_b$ increases. This `new' mode, briefly mentioned in section \ref{sec2}, will be referred to as $KH_T$ in the following, since is actually a `twisted' version (with dominant crosswise vorticity) of $KH_1$, itself the $2D$ generalization of a `classical' $1D$ $KH$ mode. Indeed, the eigenmode $KH_T$ is shown in figures \ref{fig:3Dview_KHKH2} and \ref{fig:Slice_KH_H2}, where the most unstable mode for $B=1$ is chosen. In particular, figure \ref{fig:3Dview_KH_H2} shows a qualitative $3D$ visualization of the associated perturbation velocity streamlines, together with two equal and opposite isocontours of the corresponding crosswise vorticity $\omega_z$. For comparison, figure \ref{fig:3Dview_KH} shows a similar visualization of $KH_1$, but with the `twisted' (i.e. rotated by a quarter-turn) crosswise vorticity is replaced by the (classical) spanwise vorticity $\omega_y$ of a $KH$ mode. \newline

The structure of $KH_T$ is composed of alternated counter-rotating vortices, contained in the region $-0.5 \leq z \leq 0.5$. Streamlines are looping primarily in $x-y$ planes, although they are slightly inclined along $x$. In other words, the perturbation vorticity is `twisted' so that it is primarily directed along $z$, with a slight component along $x$. This contrasts with $KH_1$ (or $H_1$), where the perturbation vorticity is primarily along $y$, which corresponds to the familiar `billow' vortices in $x-z$ planes in figure \ref{fig:3Dview_KH}. \newline

Figure \ref{fig:Slice_KH_H2} shows a more quantitative visualization of the $KH_T$ structure, with sliced view of $\omega_z$ and velocities on three orthogonal planes ($y=0, z=0, x=7.3$ respectively in the left, middle, and right column).
From the first column of figure \ref{fig:Slice_KH_H2}, it appears that isocontours of $\omega_z(y=0)$ are fully contained in the region $-0.5 \leq z \leq 0.5$, where $\partial_z U \leq 0$. Moreover, $\omega_z(y=0)$ reaches a maximum amplitude exactly at $z=0$, just as $\partial_z U(y=0) = \pi \cos{(\pi z)}$ does. Unsurprisingly, this corresponds to the $x$-location where $v(y=0)=0$, i.e the center of the vortex. From the view of $\omega_z(z=0)$ in the  $x-y$ plane, we learn that these vortices are associated with strong values of $\omega_z(z=0)$ of opposite sign near the sidewalls. This is a consequence of the no-slip boundary conditions, which also induces a strong viscous dissipation. The main vortices are slightly positively sloped along $x$, and we observe a weak but nonzero value of $w(z=0)$ perfectly in phase with $u(z=0)$. \newline

As we see in figures \ref{fig:3Dview_KHKH2} and \ref{fig:Slice_KH_H2}, contrary to the stratified case in figure \ref{fig:Harmomics}, $KH_T$ is \textit{not} an harmonic of $KH_1$ in the sense that the wavelength of $v$ of the former has decreased with respect to the one of the latter. Because the density interface does not exist at $Ri_b=0$, the spanwise velocity has no need to adapt to the increase in the wavelength of the density perturbation. In turn, the quantization proposed in Eq.(\ref{eq:lv}) does not hold in this unstratified case,  or indeed even in related relatively weakly stratified cases. This wavelength decrease in $v$ -although not in $u$ or in $w$- implies  that $KH_T$ may even become more unstable than $KH_1$ itself in  weakly stratified situations. \newline

\begin{figure}
\centering
  \begin{subfigure}[b]{0.48\linewidth}
    \includegraphics[trim={3.5cm 10cm 3.5cm 10cm},clip,width=\linewidth]{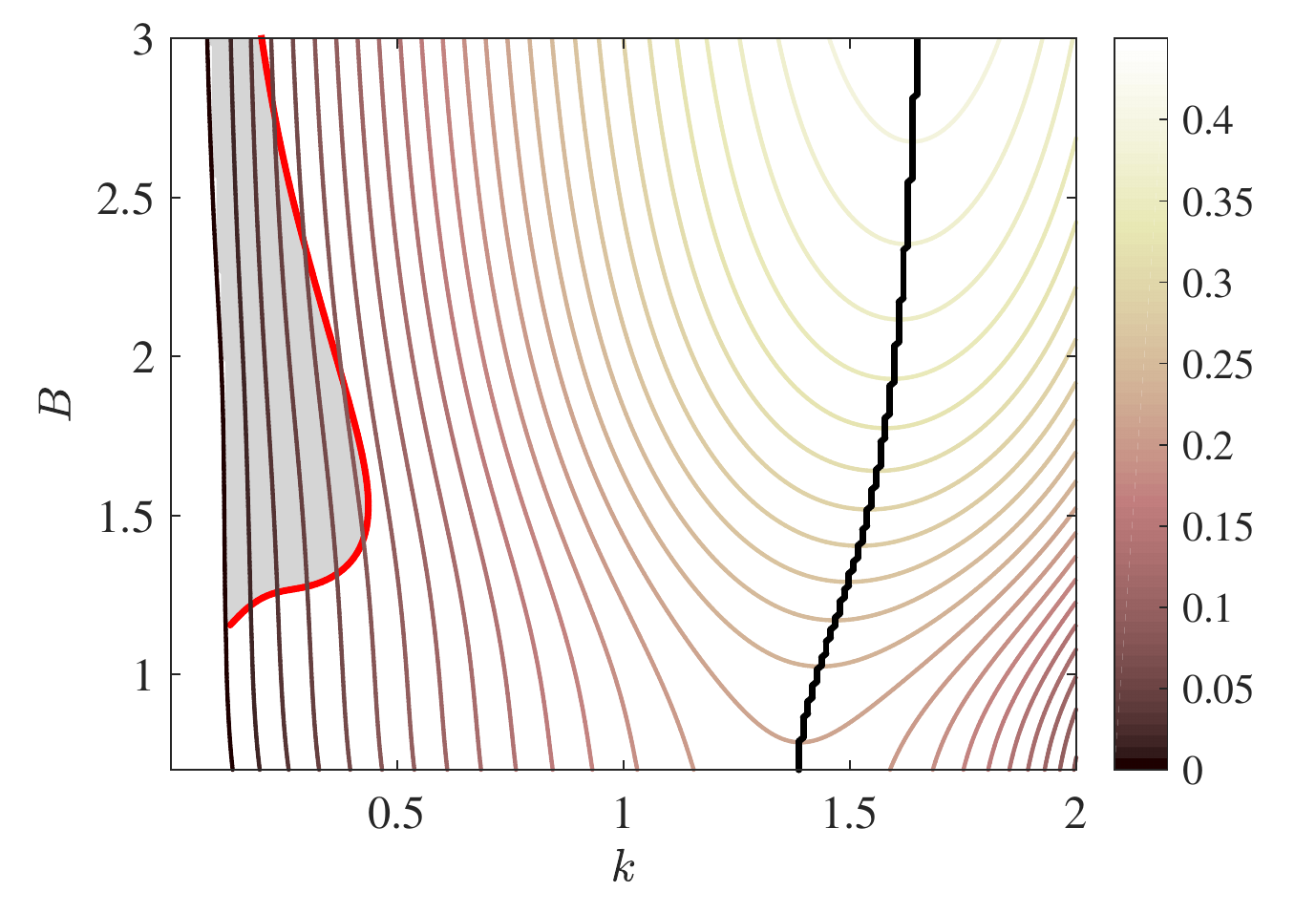}
     \caption{ $KH_1$, $Re=440$, tanh with $\gamma=5$ \label{fig:KH_H1_tanh} }
  \end{subfigure}
  \begin{subfigure}[b]{0.48\linewidth}
    \includegraphics[trim={3.5cm 10cm 3.5cm 10cm},clip,width=\linewidth]{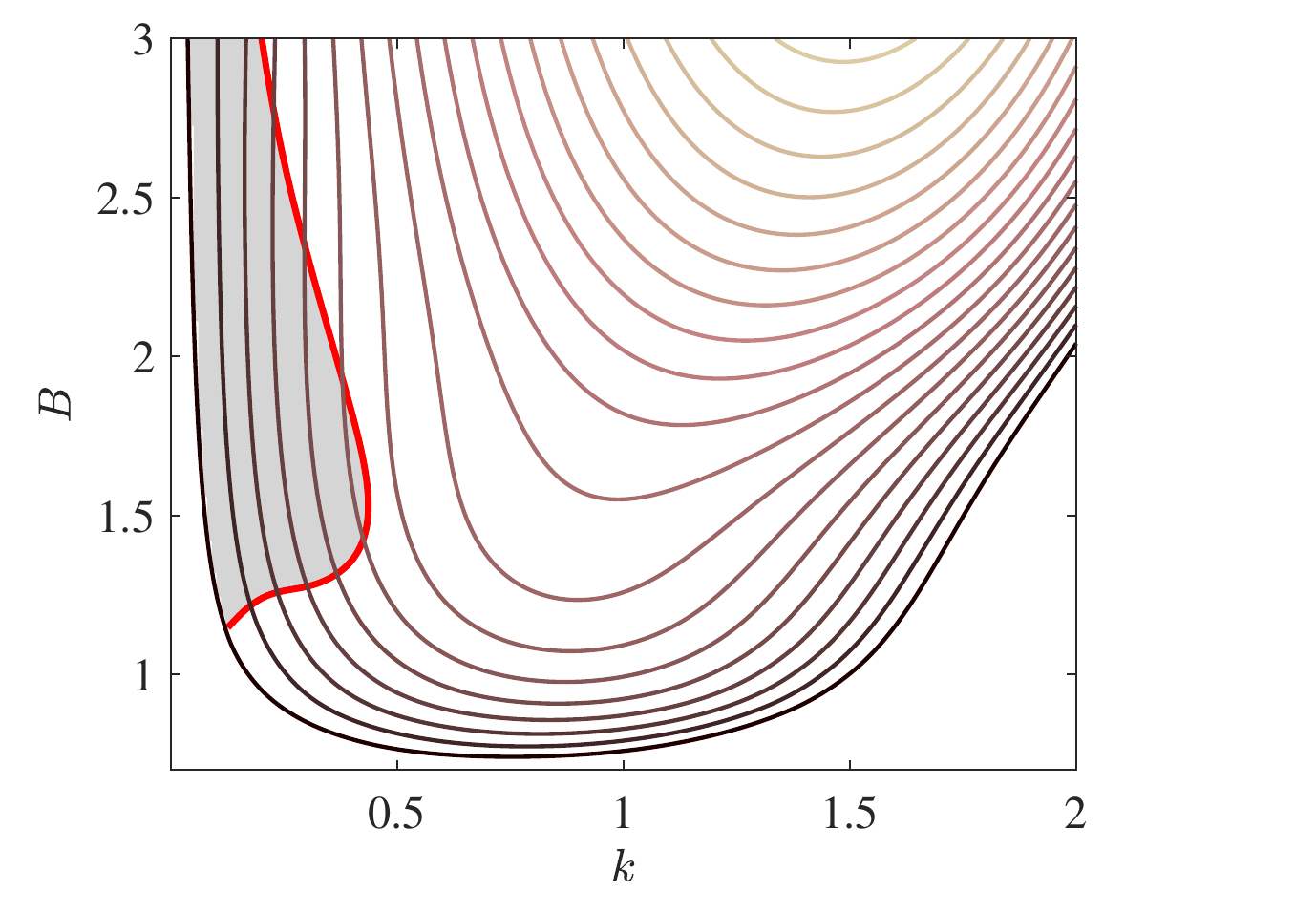}
    \caption{ $KH_T$, $Re=440$, tanh with $\gamma=5$ \label{fig:KH_H2_tanh} }
  \end{subfigure}
  \begin{subfigure}[b]{0.48\linewidth}
    \includegraphics[trim={3.5cm 10cm 3.5cm 10cm},clip,width=\linewidth]{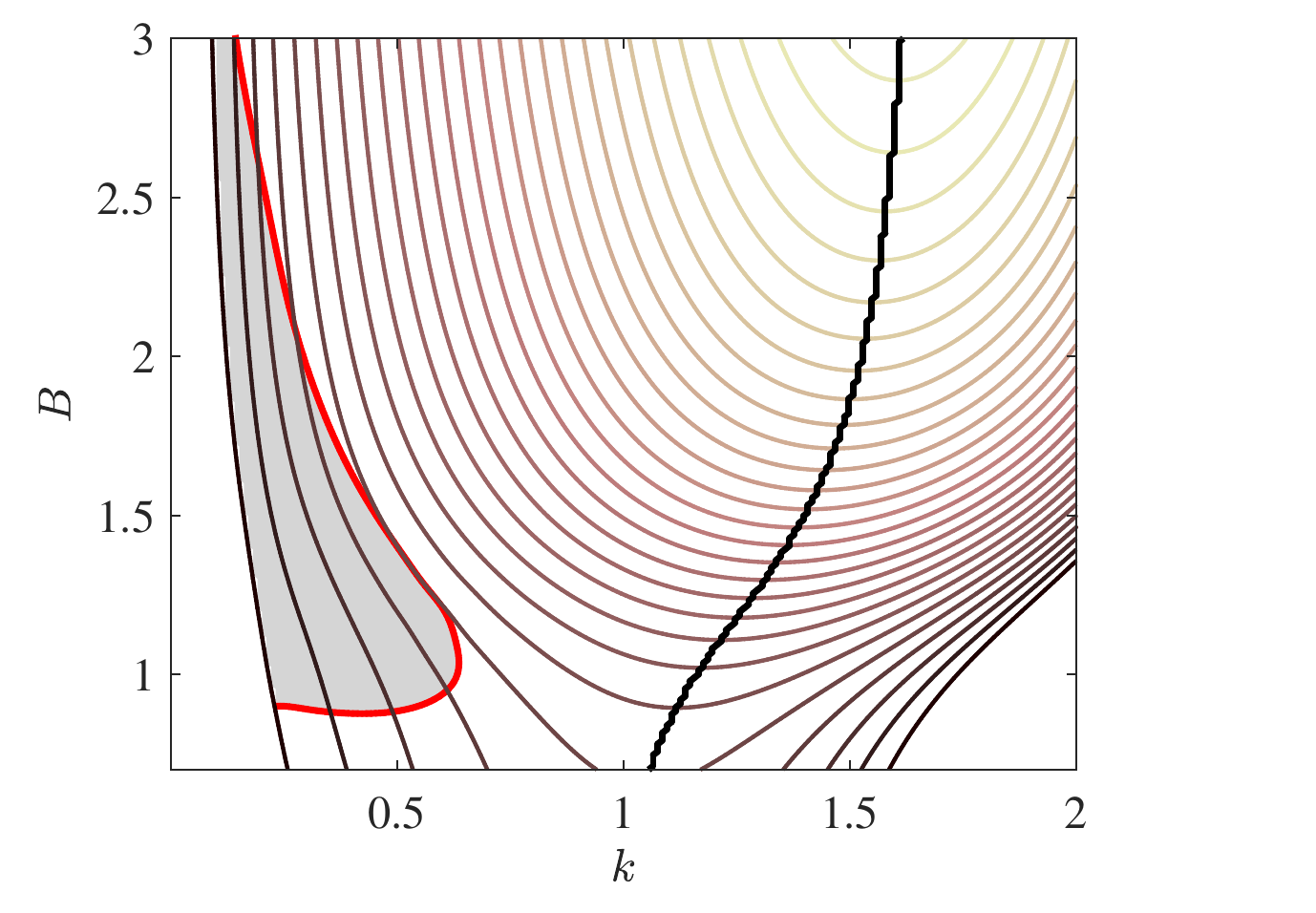}
    \caption{ $KH_1$, $Re=440$, Poiseuille \label{fig:KH_H1} }
  \end{subfigure}
  \begin{subfigure}[b]{0.48\linewidth}
    \includegraphics[trim={3.5cm 10cm 3.5cm 10cm},clip,width=\linewidth]{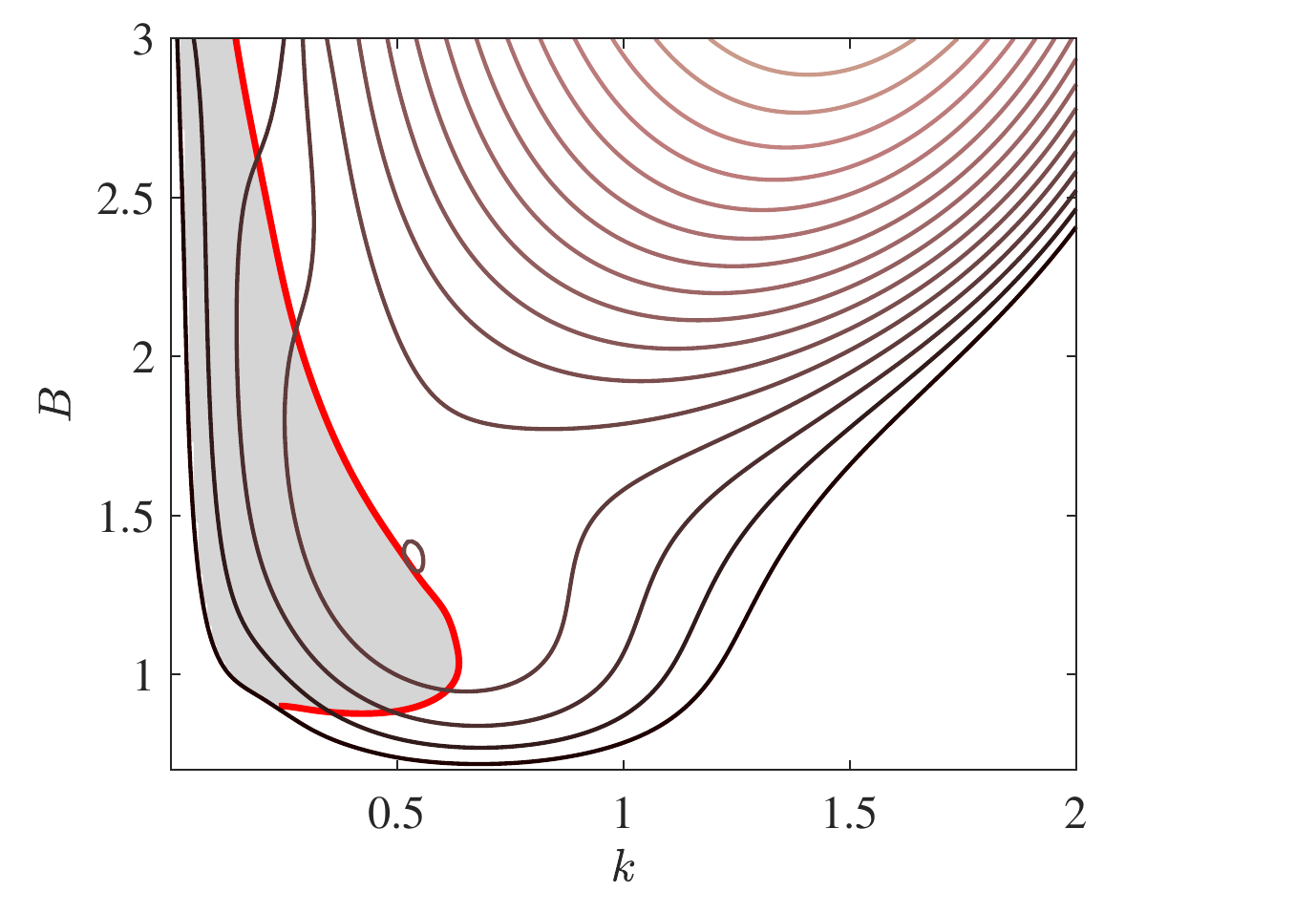}
    \caption{ $KH_T$, $Re=440$, Poiseuille \label{fig:KH_H2} }
  \end{subfigure}
  \caption{Growth rates $\sigma_r$ of $KH_1$ (left column) and $KH_T$ (right column) in the $(k,B)$ plane for $Re=440, Ri_b=0$. We compare $M_5(y)$ (top row) to $M_p(y)$ (bottom row). At the left of the red line, inside the gray-shaded area $KH_T$ is more unstable than $KH_1$. The black line is the most unstable growth rate  $ \sigma_m(B)$ over $k$.   }
   \label{fig:KH_H_GR}
\end{figure}

The growth rates of $KH_1$ and $KH_T$ are compared in the $(k,B)$ plane in figure \ref{fig:KH_H_GR} (for $Re=440, Ri_b=0$). The left and right columns show the  growth rates of $KH_1$ and $KH_T$ respectively. In addition, we investigate the effect of $M(y)$ on the stability properties of $KH_T$: the first row shows $M(y)=M_5(y)$ while the second row shows $M(y)=M_p(y)$. The red line is the locus where the growth rates of $KH_1$ and $KH_T$ are equal; on its left, in the gray-shaded area, $KH_T$ is indeed more unstable than $KH_1$. 
The black line is the most unstable growth rate $\sigma_m(B)$ over $k$, and shows that for a given aspect ratio $B$, the most unstable mode is always $KH_1$ (the red curve never crosses the black curve), as expected. However there is a low-$k$ range starting from $k=0$ where $KH_T$ locally dominates. Interestingly, the width of this range has a nonmonotonic evolution: it reaches a maximum for $B \approx 1$ for the Poiseuille base flow and $B \approx 1.5$ for the tanh base flow. Increasing $B$ above this value quickly reduces the region of dominance of $KH_T$. In other word,  two effects compete as $B$ decreases:
\begin{enumerate}
\item The first effect is that $KH_T$ `takes advantage' of confinement more than $KH_1$. Comparing figures \ref{fig:KH_H1}-\ref{fig:KH_H2} we see that as $B$  decreases, $KH_T$ has a region below $B \approx 1.75$ where $\sigma_r$ increases again. This leads to the creation of an `island' in the $\sigma_r$ contours centered around $B \approx 1.4$ and $k \approx 0.5$. In the meantime, $KH_1$ is monotonically damped, and it is precisely this divergence in the behavior of both instabilities that leads to the enlargement of the gray-shaded region where $KH_T$ dominates. The tanh profile case behaves in a similar fashion, although the span of the $KH_T$-dominated region is reduced for intermediate values of $B$. This suggests that  $KH_T$ can take advantage of the spanwise confinement only if the induced crosswise vorticity is located at the center of the duct (rather than at the walls) ; at least as long as there is enough vorticity at the center.
\item The second effect is that $KH_T$ appears much more sensitive to viscous damping (in the sense that was defined more precisely in section \ref{sec:stab}) than $KH_1$ does. As $B$ further decreases, viscous damping becomes more severe, and $KH_T$ is stabilized at a threshold $B$ where $KH_1$ is still significantly unstable. The $KH_T$-dominated region is thus shut at $B \approx 0.7-1.1$ (at these values of $Re$ and $Ri_b$).
\end{enumerate}
From these observations, the $KH_T$ mode appears to be a very delicate instability, in particular because of its ambiguous relation to spanwise confinement. A sufficiently strong confinement can `feed' $KH_T$ (on the condition that $Re$ is not too low) but not without limit: eventually an exceedingly strong confinement can suppress it (due to viscous effects). \newline


\subsection{Increased energy extraction by spanwise confinement}
 
The above observations can be brought together to build an \textit{a posteriori} explanation for the $KH_T$ instability mechanism. Proceeding as in \cite{manneville98} Chap. 7, we define the perturbation kinetic energy as:
\begin{equation}
K_p = \frac{1}{2} \int_{S} \left(\overline{u^2} + \overline{v^2} + \overline{w^2} \right)  dS  ,
\end{equation}
where the overbar denotes spatial averaging along $x$ and over an instability period, and  $\int_{S} dS = \int_{-1}^{1}\int_{-B}^{B} dz dy $ the integral over the duct cross section. By manipulating the Navier-Stokes equation and using the boundary conditions, we obtain an evolution equation for the perturbation kinetic energy :
\begin{equation}
\frac{\mathrm{d} K_p}{\mathrm{d} t} = - \int_{S} \overline{uw} \frac{\partial U}{\partial z}dS - \int_{S} \overline{uv} \frac{\partial U}{\partial y}dS - \frac{1}{Re} \int_{S}  \overline{\left \| \omega \right \|^2}  dS .  
\label{eq:per_kin}
\end{equation}
The last term in $1/Re$ stands for the viscous dissipation and can only induce loss in energy (which does not mean that viscosity has a dissipative effect \textit{only}, since it also affects $u,v,w$). The first two terms represent the rate of energy transfer, from the $z$ and $y$ mean shear to the perturbations; they may be positive and thus feed the instability. Given two complex fields $a(y,z) \doteq \left | \hat{a}(y,z) \right | e^{i\psi_a(y,z)} $ and $b \doteq \left | \hat{b}(y,z) \right | e^{i\psi_b(y,z)}$ we can write: 
\begin{equation}
\overline{ab} = \frac{1}{2}\left | \hat{a}(y,z) \right |\left | \hat{b}(y,z) \right |\cos\left[\phi_a(y,z)-\phi_b(y,z) \right ]e^{2\sigma_rt}.
\label{eq:x_av}
\end{equation}
For $t=0$, we plot in figure \ref{fig:E_terms} the two different fields $-\overline{uw}   \partial U / \partial z$ and $-\overline{uv}  \partial U / \partial y$ computed by Eq.\ref{eq:x_av} :
\begin{figure}
\centering
  \begin{subfigure}[b]{0.48\linewidth}
    \includegraphics[trim={3.75cm 9cm 4.5cm 9cm},clip,width=0.95\linewidth]{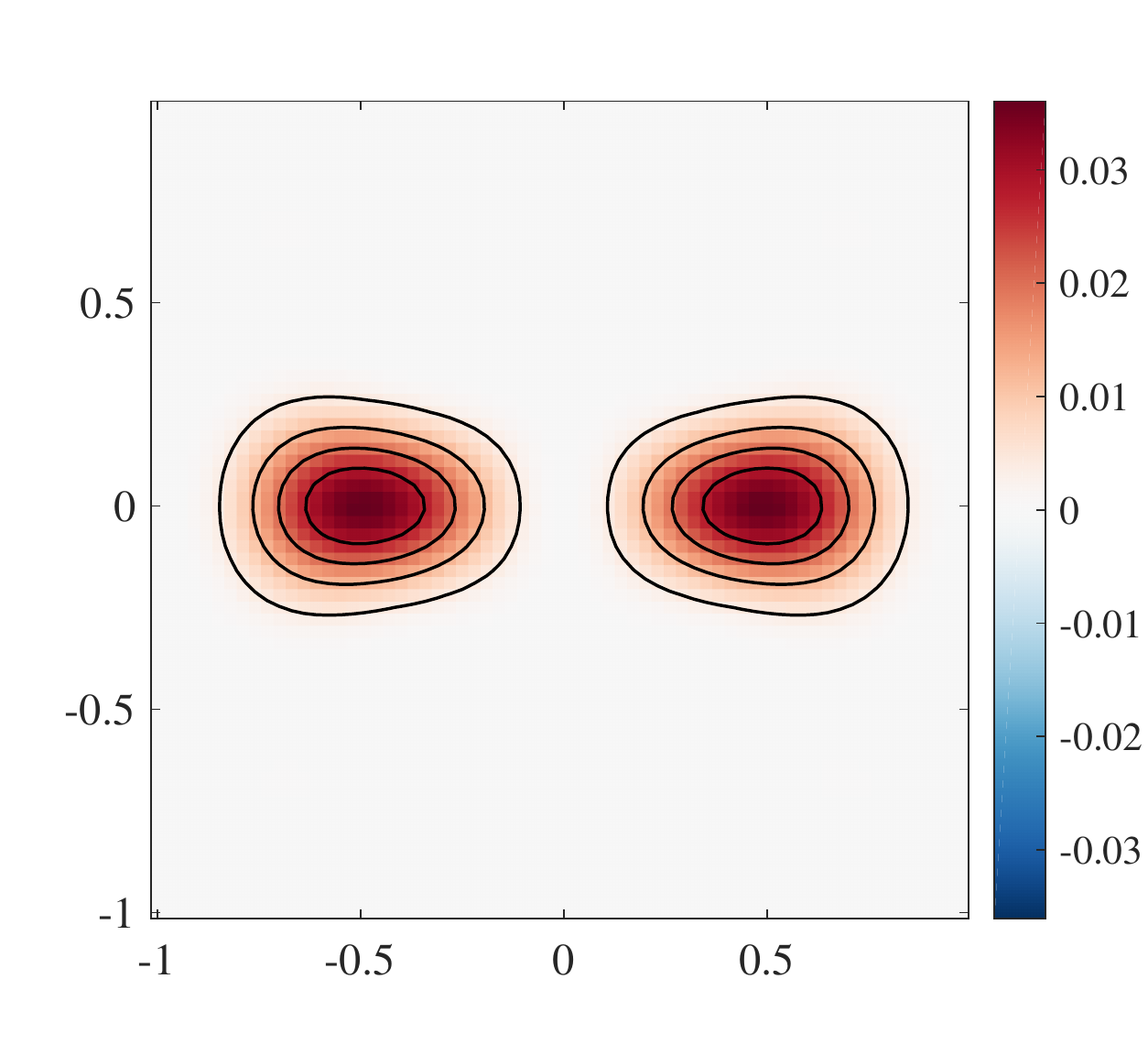}
     \caption{$-\overline{uw} \partial U / \partial z$  \label{fig:E_uwdUdz} }
  \end{subfigure}
  \begin{subfigure}[b]{0.48\linewidth}
    \includegraphics[trim={3.75cm 9cm 4.5cm 9cm},clip,width=0.95\linewidth]{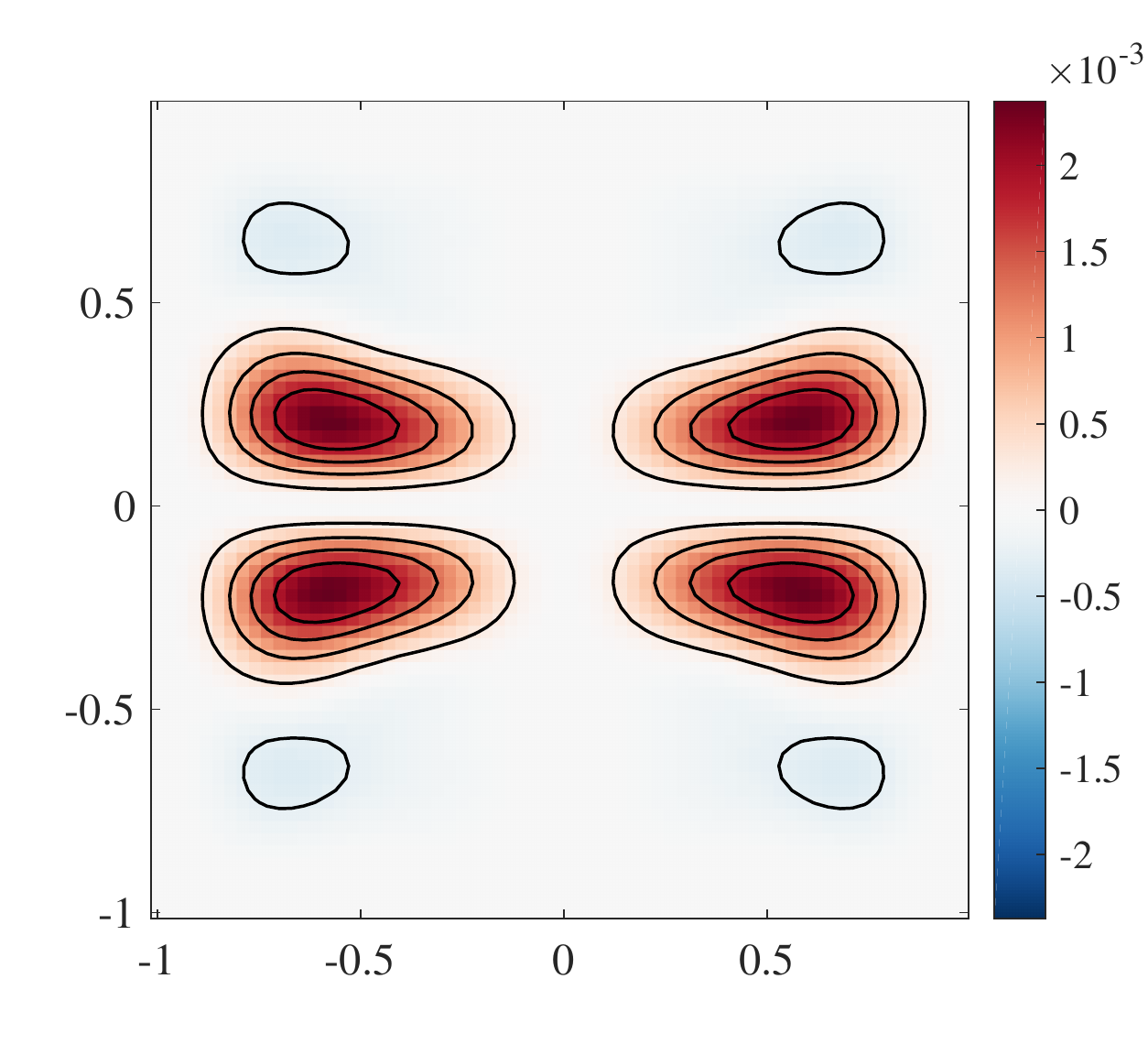}
    \caption{$-\overline{uv} \partial U / \partial y$ \label{fig:E_uvdUdy} }
  \end{subfigure}
    \caption{Energy contribution to the $KH_T$ mode shown in figure \ref{fig:3Dview_KHKH2}}
    \label{fig:E_terms}
\end{figure}
Comparing figures \ref{fig:E_uwdUdz}-\ref{fig:E_uvdUdy} immediately reveals that both mechanisms contribute to the instability (since they both promote $dK_p/dt > 0$, in red). However, the dominance of the term $-\overline{uw} \partial U / \partial z$ is evident from the scale of the colorbar. After performing the cross section integration,  the term in $ \partial U / \partial z$ in Eq. (\ref{eq:per_kin}) is approximately $12$ times larger than the one in $ \partial U / \partial y$, and is thus primarily responsible for the growth of the instability. The interesting physical implication is that, even if the structure of the $KH_T$ mode differs completely from the $KH_1$ and 1D $KH$ modes (since its  vorticity is principally along $z$ and not along $y$), it is equally fed by the \textit{spanwise} vorticity of the base flow $\partial U / \partial z$. The positiveness of [$- \int_{S} \overline{uw}  \partial U / \partial z dS$] is ensured by that fact that $\overline{uw} > 0$ in the whole cross section and $\partial_z U <0$ such that the main shear transfers energy into the perturbation vortices. This is a consequence of the fact that $u$ and $w$, concentrated in the region $-0.5 \leq u \leq 0.5$, are almost perfectly in-phase. This also means that perturbation vortices are slightly positively sloped along $x$, and thus feel the main shear $\partial_z U$. \newline

These results  suggest that for a very weak stratification ($Ri_b \approx 0$), and strong confinement (small $B$), a low wavenumber mode (small $k$) that is intrinsically $2D$ can unexpectedly become more unstable than the quasi-$1D$ $KH_1$ wave. This `twisted' $KH_T$ mode may be observed in practice, but only if long waves are preferentially forced. 
\section{Conclusions \label{sec:conc}}
In this paper we have compared confined $2D$ dispersion relations with $1D$ spanwise invariant ones, for different aspect ratio $B$ and bulk Richardson numbers $Ri_b$ characterizing the stratified sheared flows in a rectangular duct. In this limited parameter space, the presence of duct walls has a stabilizing effect except in a negligibly small region. Namely, the $1D$ predictions are almost systematically an upper bound for the $2D$ growth rates, which decrease monotonically as the lateral walls tighten around the flow. The natural question that arises thus concerns the threshold aspect ratio above which the $1D$ prediction is  relevant. We have shown that the answer is sensitive to the precise structure of the base flow: the thinner the spanwise boundary layers, the lower this minimal aspect ratio. Although less significant, the same conclusion regarding the influence of increasing $Ri_b$ can also be drawn. \newline
We have largely  restricted ourselves to a particular $Ri_b$ and wavenumber, allowing for a more detailed analysis of the unstable part of the spectrum. In the $2D$ case, a plethora of unstable modes is observed, and they proliferate as the aspect ratio is increased. These modes present a regularity in their spatial structures, which is perhaps not really all that  surprising. Furthermore, from moderate to infinite aspect ratio, the $1D-O$ dispersion relation for oblique waves seems to be very close to a continuous version of the $2D$ spectrum, provided the base flow spanwise boundary layer is sufficiently thin. This link is made clear by noticing that, thanks to the spanwise symmetry of the system, the $1D-O$ dispersion relation also incorporates modes that mimic the same structural regularity as the $2D$ modes. Here,  the quality of the $2D$/$1D-O$ alignment is slowly improved by increasing the aspect ratio, and quickly improved by thinning the spanwise boundary layers of the base flow. Implications of these observations are believed to be important. As the confinement widens, the most unstable $2D$ modes are competing more and more tightly: a slight external forcing on one of their particular wavelength is sufficient to make it emerge preferentially with respect to the one predicted by the $1D$ analysis. Thus, even in the large aspect ratio limit, the $1D$ predictions must be taken with a lot of caution, and should be complemented by a $1D-O$ analysis.\newline
In the $2D$ context, we expect the mode that oscillates the least in the spanwise direction to be the most unstable one, as a consequence of viscous damping. We have finally shown that, for a very restricted range of parameters,  a mode whose spatial structure resembles a 'twisted' version of $KH_1$ (in terms of having dominant crosswise vorticity) becomes more unstable than the classical $KH_1$ mode. This phenomenon has no $1D$ counterpart, and is shown to be inherently linked to a tight confinement from which this mode takes advantage.  \newline 
Looking ahead, these results may have interesting implications that could stimulate future research. In line with qualitative comments made in section \ref{sec1}, the evolution of the group velocities deduced from figure \ref{fig:2D_DR} suggests a potential convective-to-absolute transition occurring as the lateral walls are brought closer together. A rigorous saddle point or impulse response approach would be needed to shed light on the existence of this transition. Such an analysis appears of particular relevance, as in practice the duct is of finite length in the streamwise direction. Thus, it remains unclear if the structure in \citep{Lefauve18} is the product of convective instabilities reflected at the extremities of the duct connecting with the reservoirs, or if it would remain self-sustained in an infinite domain. Confinement may play a crucial role in such discrimination, as shown by numerous examples in the literature for unstratified flows (for example \citep{Juniper06}, \citep{Healey09}, \citep{Rees10}, among others). \newline 
In the $2D$ case, the presence of modes of comparable growth rates could lead to a very rich nonlinear dynamics. This is particularly true considering that these modes are naturally structural harmonics of each others. Thus, the nonlinear creation of higher harmonics of the most unstable one may be strongly amplified and lead to powerful interactions. 

\begin{acknowledgments}

\end{acknowledgments}

\appendix
\section{Formulation of $1D$ linear stability problem}
\label{1D_pb}
The one-dimensional stability problem reduces to solving (for $w$ and $\rho$):
\begin{equation}
\setlength{\arraycolsep}{1pt}
\renewcommand{\arraystretch}{1}
\sigma\left[
\begin{array}{cc}
  \Delta  \ \ &     \\
    &  \mathcal{I}  \\  
\end{array}  \right] 
\setlength{\arraycolsep}{1pt}
\renewcommand{\arraystretch}{1}
\left[
\begin{array}{c}
  w \\
  \rho\\  
\end{array}  \right]=
\setlength{\arraycolsep}{1pt}
\renewcommand{\arraystretch}{1}
\left[
\begin{array}{cc}
  \mathcal{L}_w  \ \ &  \mathcal{L}_{w\rho}   \\
\mathcal{L}_{\rho w}    &  \mathcal{L}_{\rho}  \\  
\end{array}  \right]
\setlength{\arraycolsep}{1pt}
\renewcommand{\arraystretch}{1}
\left[
\begin{array}{c}
  w \\
  \rho\\  
\end{array}  \right] ,
\label{GenEig}
\end{equation}
where
\begin{subeqnarray*}
\mathcal{L}_w & = & -\mbox{i}kU\Delta + \mbox{i}k\partial_{zz}U + Re^{-1}\Delta^2 ,\\
\mathcal{L}_{\rho w} & = & -\mbox{i}kU + (Re Sc)^{-1}\Delta ,\\
 \mathcal{L}_{w \rho} & = & Ri_b\left[ k^2 \cos{\theta} - \mbox{i} k \sin{\theta}\partial_z \right] \label{eq:Lrw} ,\\ 
\mathcal{L}_{\rho} & = & - \partial_z R ,
\end{subeqnarray*}
with
\begin{equation}
U(z) = - \sin{(\pi z)}, \ \ \ \ \ \ \ \ \ \ \ -1 \leq z \leq 1,
\label{eq:U}
\end{equation}
 and
\begin{equation}
\mathcal{R}(z) = - \tanh\left(2 R (z-z_0) \right)\ , \ \ \ \ \ \ \ \ \ \ \ -1 \leq z \leq 1 .
\end{equation}
\section{Formulation of $1D$-$O$blique linear stability problem}
\label{1Dobl_pb}
This problem corresponds to Eq.(\ref{GenEig_2D}), after the $y$ dependence of the base flow is removed, and after the $\partial_y$ acting on the perturbations are replaced by $i \beta$. After factorization by $v$,$w$,$\rho$ and $p$, we end up with the system:  
\begin{align}
\begin{split}
\left[ \sigma + Uik -Re^{-1}(-k^2+\partial_{zz} - \beta^2) \right]v  &= - i\beta p  ,\\
\left[\sigma + Uik -Re^{-1}(-k^2+\partial_{zz} - \beta^2) \right]w &= - \partial_z p - Ri\cos(\theta)\rho   ,\\
\left[\sigma \rho + Uik - (Sc Re)^{-1}(-k^2+\partial_{zz} - \beta^2) \right] \rho & = -  (\partial_z R) w  ,\\
\left[ \sigma i\beta  - kU\beta - Re^{-1}( -k^2i\beta -i \beta^3  + i \beta\partial_{zz} ) \right]v  &= \\
\left[ -\sigma\partial_z -ikU \partial_z +ik\partial_z U + Re^{-1}(-k^2 \partial_z -\beta^2\partial_{z}  + \partial_{zzz} ) \right] w  & - k^2 p  - ik Ri_b \sin(\theta)\rho .
\end{split}
\end{align}
$v$ is always multiplied by an odd power of $\beta$, when, in the meantime, $\rho$, $w$, $p$ and $u$ (by continuity) are multiplied by an even power of $\beta$. Thus, we easily show that the system is invariant under the transformation 
\begin{equation}
(\sigma, \beta, u, v, w, \rho) \rightarrow (\sigma , - \beta, C u, -C v , C u,  C w ,  C \rho ),
\end{equation}
where $ C \in \mathbb{C}$, a constant, is an unimportant consequence of the linearity of the system. Interestingly, $v$ must flip its sign with respect to all the others fields as $\beta \rightarrow - \beta$. The existence of this symmetry has an important consequence. Indeed, by linearity of the system, perturbations of the form 
\begin{align}
\begin{split}
\hat{h}(y,z) = \overline{h}(z)e^{i\beta y} + \overline{h}(z)e^{-i\beta y} \doteq \breve{h}(z)\cos(\beta y) , \\
\hat{v}(y,z) = \overline{v}(z)e^{i\beta y} - \overline{v}(z)e^{-i\beta y} \doteq \breve{v}(z)\sin(\beta y) ,
\end{split}
\end{align}
and perturbations of the form
\begin{align}
\begin{split}
\hat{h}(y,z) = \overline{h}(z)e^{i\beta y} - \overline{h}(z)e^{-i\beta y} \doteq  \breve{h}(z)\sin(\beta y) ,\\
\hat{v}(y,z) = \overline{v}(z)e^{i\beta y} + \overline{v}(z)e^{-i\beta y} \doteq  \breve{v}(z)\cos(\beta y) , \\
\end{split}
\end{align}
\noindent are equally solutions (where $\overline{h}$ denotes any one of $u$, $w$, $\rho$ or $p$ ). In other words, they satisfy the $1D-O$ dispersion relation. This may appear surprising because, although these solutions are periodic in the spanwise direction, they travel purely along the streamwise direction (standing wave), rather than obliquely. We notice that $\breve{v}$ is necessarily phase-shifted by $i=\sqrt{-1}$ with respect to all the other fields. These solutions could be made more general by adding the same constant phase shift in the harmonic function of $\hat{h}(y,z)$ and $\hat{v}(y,z)$. 
\section{Numerical methods \label{appnm}}
In the following, we present the numerical methods used for the discretization and solution of the system in Eq.(\ref{GenEig_2D}). Since our study required us to perform a much greater number of computations than in \cite{Lefauve18}, we developed a more accurate and faster method than their finite-difference method.
\subsection{Discretizing the equations}
\paragraph*{Expansions --} Eq.(\ref{GenEig_2D}) is discretized by a purpose-built pseudospectral Chebyshev method, together with a crosswise mapping concentrating points at the density interface. The boundary conditions were built directly into the basis functions, so that the eigenfunctions necessarily satisfy the boundary conditions. The perturbation fields are expanded as:
\begin{subeqnarray}
v(y,z) &=& \sum_{m=0}^{N_y-1}\sum_{n=0}^{N_z-1} V_{mn}  \phi_m(s_y[y]) \zeta_n(s_z[z]) ,\\
w(y,z) & = & \sum_{m=0}^{N_y-1}\sum_{n=0}^{N_z-1} W_{mn}  \zeta_m(s_y[y]) \phi_n(s_z[z]), \\
\rho(y,z) & = & \sum_{m=0}^{N_y-1}\sum_{n=0}^{N_z-1} R_{mn}  \psi_m(s_y[y]) \psi_n(s_z[z]) ,\\
p(y,z) & = & \sum_{m=0}^{N_y-1}\sum_{n=0}^{N_z-1} P_{mn}  T_m(s_y[y])  T_n(s_z[z]) ,
\end{subeqnarray}
where $T_j(x)$ are the Chebyshev polynomial of order $j$.\newline 
\paragraph*{Basis functions --} The three sets of functions $\zeta_j(x)$, $\psi_j(x)$ and $\phi_j(x)$  respectively satisfy Dirichlet, Neumann and Dirichlet-Neumann boundary conditions at $x = \pm 1, \forall j$. Classically, they are well-chosen linear combinations of Chebyshev polynomials. Expressions for such functions were proposed, e.g. in \cite{Kato00}, which have been checked to be suitable in the present case too. However, a well-known disadvantage of Chebyshev polynomials is their intense (and high-frequency) oscillations near boundaries, dangerously blowing up with the order of differentiation: $\left | \mathrm{d}^p T_N(\pm 1)/\mathrm{d} x^p \right | \sim N^{2p}$ \cite{Boyd00}. As the present problem contains derivatives up to third order (trough the terms $\partial_{yyy}v$ and $\partial_{xxx}w$), the corresponding discretization matrices may inherit a $O(N^6)$ condition number, which could become particularly challenging for fine grids. To overcome this problem, we adopt the method proposed by Heinrichs \cite{Heinrich89}:    
\begin{subeqnarray*}
\zeta_j(x) &=& (1-x^2)  T_j(x) , \\
\phi_j(x) &=& (1-x^2)^2  T_j(x) ,
\end{subeqnarray*}
where we easily check that $\zeta(\pm 1) = \phi(\pm 1) = \phi^{'}(\pm 1) =0$. In the $p$-th derivative of $\zeta_j(x)$, the prefactor $1-x^2$ kills the $\mathrm{d}^p T_j/\mathrm{d} x^p$ term at $x=\pm 1$, leading to a new condition number $O(N^{2(p-1)})$. The same applies to $\phi_j(x),\mathrm{d}^p T_j/\mathrm{d} x^p, \mathrm{d}^{p-1} T_j/\mathrm{d} x^{p-1}$, leading to a condition number $O(N^{2(p-2)})$. Therefore, our discretization matrices are at worst $O(N^2)$ for both $\zeta$'s (up to second-order derivative) and $\phi$'s (up to third order one). For $\psi$'s, which are only used in the density perturbation expansion, we adopted the expression proposed by \cite{Boyd00}:
\begin{subeqnarray*}
\psi_{2n}(x) &=& 
\left\{
    \begin{array}{ll}
     1 &  n = 0 \\[2pt]
      T_{2n}(x) - \left [\frac{n^2}{(n+1)^2}  \right ]T_{2n+2}(x) &  n=1,2,...
  \end{array} \right. \\
\psi_{2n+1}(x) &=& T_{2n+1}(x) - \left [\frac{2n+1}{2n+3}  \right ]^2T_{2n+3}(x) \ \ \ \ n=0,1,...
\end{subeqnarray*}
where $\psi^{'}(\pm 1) = 0$. \newline

\paragraph*{Collocation points --}  We chose `Gauss-Chebyshev' collocation points, equivalent to the roots of the Chebyshev polynomials:
\begin{equation}
s_y =  \cos{\left[\frac{(2i-1)\pi}{2 N_y} \right]} \quad \quad s_z = \cos{\left[\frac{(2i-1)\pi}{2 N_z} \right]},
\label{eq:GC_points}
\end{equation}
where $s_y$ and $s_z$ designate respectively  the spanwise and crosswise (vertical) collocation points. This choice  contrasts with the classical `Gauss-Lobatto' discretization, and excludes boundary points. This is deliberate in order to avoid spurious pressure modes, inherent to Gauss-Lobatto meshing. If boundary points are needed, for instance with the use of the tau method, the $\mathbb{P}_{N}-\mathbb{P}_{N-2}$ technique presented in \cite{Peyret02} is a suitable alternative. Because it relies on the interpolation of the pressure field, it however excludes all nonlinear mapping. Indeed the Gauss-Lobatto points locations, optimal for a polynomial interpolation, are then distorted and the Runge phenomena is observed at the boundaries, precisely where we desire the pressure values. \newline

\paragraph*{Coordinate mapping --}  Under mapping transformations, the physical points corresponding to the numerical grid of Eq.(\ref{eq:GC_points}) are recovered as:
\begin{subeqnarray*}
y &=& g_y(s_y) = A  s_y , \\
z &=& g_z(s_z ; \alpha_1 , \alpha_2) = \alpha_2  + \frac{\tan\left[(s_z-s_0)\lambda \right]}{\alpha_1} 
\end{subeqnarray*}
where 
\begin{subeqnarray*}
s_0 = \frac{\kappa -1 }{\kappa + 1}, \quad
 \kappa = \frac{\arctan(\alpha_1(1+\alpha_2))}{\arctan(\alpha_1(1-\alpha_2))}, \quad
 \lambda = \frac{\arctan(\alpha_1(1-\alpha_2))}{(1-s_0)},
\end{subeqnarray*}
The simple linear mapping $g_y$ transforms the interval $y \in [-A,A]$ into $s_y \in [-1,1]$, where the Chebyshev polynomials are defined.  The mapping $g_z$ (see \cite{Bayliss92}), concentrates the collocation points around $z=\alpha_2$ with a strength modulated by $\alpha_1$. As a sharp evolution of the density perturbation is expected at the interface, we set $\alpha_2 = z_0$. An optimum for $\alpha_1$ can be found trough the technique proposed in \cite{Bayliss92}. However, the present problem is slightly different since the mapping also impacts on the velocity: a compromise was found by trial and error to obtain an $\alpha_1$ sufficiently big as to smooth out the density, but sufficiently small as not to  distort the velocity substantially. Overall, this crosswise mapping  sped up the convergence impressively.  \newline

As mentioned in \cite{Boyd00}, despite the use of mappings,  the whole problem can still be solved numerically solely in terms of physical variables and grid. In the code, this  requires two additional subroutines that: (i) computes the physical points from \ref{eq:GC_points}; (ii) performs the chain rule to transform $s$-derivatives into $z$ ones. These pivoting expressions are analytically expressed as:
\begin{subeqnarray*}
\label{chainderi}
\frac{\mathrm{d} }{\mathrm{d} z} &=& \left(\frac{1}{g_1}\right)\frac{\mathrm{d} }{\mathrm{d} s} ,\\
\frac{\mathrm{d}^2 }{\mathrm{d} z^2} &=& \frac{1}{g_1^3}\left(g_1\frac{\mathrm{d}^2 }{\mathrm{d} s^2} - g_2\frac{\mathrm{d} }{\mathrm{d} s}\right) ,\\
\frac{\mathrm{d}^3 }{\mathrm{d} z^3} &=& \frac{1}{g_1^5}\left[ g_1^2\frac{\mathrm{d}^3 }{\mathrm{d} s^3} - 3g_1g_2\frac{\mathrm{d}^2 }{\mathrm{d} s^2} + \left(-g_3g_1+3g_2^2 \right)\frac{\mathrm{d} }{\mathrm{d} s}\right] , \\
\frac{\mathrm{d}^4 }{\mathrm{d} z^4} &=& \frac{1}{g_1^7}\left[ g_1^3\frac{\mathrm{d}^4 }{\mathrm{d} s^4} -6 g_2g_1^2\frac{\mathrm{d}^3 }{\mathrm{d} s^3} + \left(-4g_3g_1^2 + 15g_2^2g_1\right)\frac{\mathrm{d}^2 }{\mathrm{d} s^2} + \left(-g_4g_1^2 + 10 g_3g_2g_1 - 15g_2^3 \right)\frac{\mathrm{d} }{\mathrm{d} s}\right] ,\\
\end{subeqnarray*}
where we use the following shorthand notation for derivatives $g_1=g^{'}(s)$, $g_2=g^{''}(s)$, $g_3=g^{'''}(s)$ and $g_4=g^{''''}(s)$.  For the $s$-derivatives, the chain rule must also be used to transform the $T$ derivatives into $\zeta$ and $\phi$ ones  (analytical expression in \cite{Boyd00}).  \newline
In order to illustrate the discretization machinery, the $\mathcal{L}_v$ operator becomes 
\begin{equation*}
\mathbf{L}^v = -i k \mathbf{U} + Re^{-1}( -k^2 \mathbf{I}^v + \mathbf{D}_{yy}^v + \mathbf{D}_{zz}^v ) ,\end{equation*}
with
\begin{equation*}
\mathbf{I}^v = \mathbf{D}^v_{0,y} \otimes \mathbf{D}^v_{0,z},\quad  \mathbf{D}_{yy}^v = \mathbf{D}^v_{2,y}\otimes \mathbf{D}^v_{0,z},\quad   \text{and} \quad \mathbf{D}_{zz}^v  = \mathbf{D}^v_{0,y}\otimes \mathbf{D}^v_{2,z}.
\end{equation*}
In addition, we express as in a $1D$ problem:
\begin{equation*}
    [\mathbf{D}^v_{0,y}]_{ij} = \frac{\mathrm{d} \phi_j(y_i)}{\mathrm{d} y} ,\quad  [\mathbf{D}^v_{2,y}]_{ij} = \frac{\mathrm{d}^2 \phi_j(y_i)}{\mathrm{d} y^2} ,\quad  [\mathbf{D}^v_{0,z}]_{ij} = \frac{\mathrm{d} \zeta_j(z_i)}{\mathrm{d} z},\quad  \text{and}  \quad [\mathbf{D}^v_{2,z}]_{ij} =\frac{\mathrm{d}^2 \zeta_j(z_i)}{\mathrm{d} z^2}.
\end{equation*}
\newline   
Processing similarly for all operators results in a new generalized eigenvalue problem:
\begin{equation}
\sigma\mathbf{B} \boldsymbol{x} = \mathbf{A} \boldsymbol{x},
\label{GenEig_num_c1}
\end{equation}
where $\boldsymbol{x} = \left[\mathbf{V},\mathbf{W},\mathbf{R},\mathbf{P} \right]$, and where $\mathbf{A}$ and $\mathbf{B}$ are $(4  Ny Nz )\times (4  Ny  Nz) $ matrices.
\subsection{Solving the discrete system}
Provided $k$ ($\in \mathbb{R}$), and a `shift' $\mu$ ($\in \mathbb{C}$) close to where the eigenvalue is sought, the system of Eq.(\ref{GenEig_num_c1}) is solved for $\sigma$ ($\in \mathbb{C}$) using the shift and invert algorithm. The selected eigenvalue among the full spectrum is that nearest to $\mu$. The procedure detailed in \cite{Hu12} is followed, except that we preferred a QR decomposition to their LU decomposition for reasons of numerical stability. In particular, the shift and inverted matrix $\mathbf{K} = (\mathbf{A}-\mu \mathbf{B})^{-1} \mathbf{B} $ is computed as:
\begin{subeqnarray}
\mathbf{QR} &=& \mathbf{A}-\mu \mathbf{B}, \slabel{eq:qr} \\
\mathbf{C} &\doteq & \mathbf{Q}^{-1}\mathbf{B} = \mathbf{Q}^{T}\mathbf{B},\\
\mathbf{K} &=& \mathbf{R}^{-1}\mathbf{C}, \slabel{eq:rc}
\end{subeqnarray}
\noindent where the decomposition in Eq.(\ref{eq:qr}), as well as the inversion in Eq.(\ref{eq:rc}), are respectively performed by `\texttt{[Q,R] = qr(A-mu*B)}'  and `\texttt{K = R\textbackslash C}' commands in MATLAB. Still following terminology in  \cite{Hu12}, the generalized eigenvalue problem ($\ref{GenEig_num_c1}$) can be rewritten as a standard one:
\begin{equation}
\mathbf{K} \boldsymbol{x} = \theta \boldsymbol{x},
\end{equation}
where $\theta = (\sigma - \mu)^{-1}$. It is solved using the implicitly restarted Arnoldi method embedded in  MATLAB's function `\texttt{eigs}', together with the `\texttt{lm}' (largest magnitude) option. Choosing this option ensures that the selected eigenvalue maximizes the quantity $\left \|  \theta \right \| = \left \|  (\sigma - \mu)^{-1} \right \|$, thus minimizing the distance between $\sigma$ and $\mu$. \newline 
Sometimes, the full spectrum is of interest, in particular when an initial guess for $\mu$ is sought. In this case, the QZ algorithm is used directly for the eigenvalues of Eq.(\ref{GenEig_num_c1}) through MATLAB's function `\texttt{eig(A,B)}'.

\bibliography{main}

\newpage

\end{document}